\documentclass{article}

\usepackage{arxiv}

\usepackage[utf8]{inputenc} 
\usepackage[T1]{fontenc}    
\usepackage{hyperref}       
\usepackage{url}            
\usepackage{booktabs}       
\usepackage{amsfonts}       
\usepackage{nicefrac}       
\usepackage{microtype}      
\usepackage{lipsum}		    
\usepackage{graphicx}
\usepackage{doi}

\usepackage{subfig}
\usepackage{physics}
\usepackage{multirow}
\usepackage[table]{xcolor}

\definecolor{orange-red}{rgb}{1.0, 0.27, 0.0}
\definecolor{seagreen}{rgb}{0.18, 0.55, 0.34}
\definecolor{lightsalmonpink}{rgb}{1.0, 0.6, 0.6}
\definecolor{darkseagreen}{rgb}{0.56, 0.74, 0.56}
\definecolor{salmon}{rgb}{1.0, 0.55, 0.41}
\definecolor{salmonpink}{rgb}{1.0, 0.57, 0.64}

\definecolor{grannysmithapple}{rgb}{0.66, 0.89, 0.63}
\definecolor{coralpink}{rgb}{0.97, 0.51, 0.47}

\newcommand{\myo}{\cellcolor{coralpink}}
\newcommand{\myl}{\cellcolor{grannysmithapple}}
\newcommand{\myg}{\cellcolor{lightgray}}

\definecolor{colorA}{RGB}{255,  69,   0}
\definecolor{colorB}{RGB}{ 46, 139,  87}
\definecolor{colorC}{RGB}{250, 128, 114}
\definecolor{colorD}{RGB}{143, 188, 143}

\title{Protecting Feed-Forward Networks from Adversarial Attacks Using Predictive Coding}

\date{May 22, 2024}	
\date{} 			

\author{ 
\href{https://orcid.org/0009-0007-6372-3539}{\includegraphics[scale=0.06]{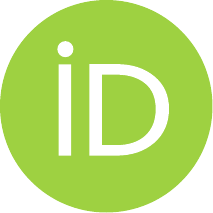}
\hspace{1mm}Ehsan Ganjidoost}\thanks{Neurocognitive Computing Lab} \\
        Cheriton School of Computer Science\\
	University of Waterloo\\
	Waterloo, ON \\
	\texttt{eganjido@uwaterloo.ca} \\
\And
\href{https://orcid.org/0000-0002-4897-8951}{\includegraphics[scale=0.06]{orcid.pdf}
\hspace{1mm}Jeff Orchard} \\
	Cheriton School of Computer Science\\
	University of Waterloo\\
	Waterloo, ON \\
	\texttt{jorchard@uwaterloo.ca} \\
}

\renewcommand{\shorttitle}{\textit{arXiv} Template}

\hypersetup{
pdftitle={Predictive Coding Survives on Perturbation Attacks},
pdfsubject={q-bio.NC, q-bio.QM},
pdfauthor={Ehsan Ganjidoost, Jeff Orchard},
pdfkeywords={First keyword, Second keyword, More},
}

\begin{document}
\maketitle

\begin{abstract}

An adversarial example is a modified input image designed to cause a Machine Learning (ML) model to make a mistake; these perturbations are often invisible or subtle to human observers and highlight vulnerabilities in a model's ability to generalize from its training data. Several adversarial attacks can create such examples, each with a different perspective, effectiveness, and perceptibility of changes. Conversely, defending against such adversarial attacks improves the robustness of ML models in image processing and other domains of deep learning.
Most defence mechanisms require either a level of model awareness, changes to the model, or access to a comprehensive set of adversarial examples during training, which is impractical. Another option is to use an auxiliary model in a preprocessing manner without changing the primary model.
This study presents a practical and effective solution -- using predictive coding networks (PCnets) as an auxiliary step for adversarial defence. By seamlessly integrating PCnets into feed-forward networks as a preprocessing step, we substantially bolster resilience to adversarial perturbations. Our experiments on MNIST and CIFAR10 demonstrate the remarkable effectiveness of PCnets in mitigating adversarial examples with about $82\%$ and $65\%$ improvements in robustness, respectively. The PCnet, trained on a small subset of the dataset, leverages its generative nature to effectively counter adversarial efforts, reverting perturbed images closer to their original forms. This innovative approach holds promise for enhancing the security and reliability of neural network classifiers in the face of the escalating threat of adversarial attacks.

\end{abstract}


\newcommand{\pcDynamics}{
    \begin{align}
        \tau\, \dot{\varepsilon}_i &= v_i - M_{i} \sigma(v_{i+1})  - b_i - \xi \varepsilon_i \label{eq:errorDynamic} \\
        \tau\, \dot{v}_i &= -\varepsilon_i + W_{i-1} \varepsilon_{i-1} \odot \sigma'(v_{i}) \label{eq:valueDynamic}\\
        \gamma\, \dot{M}_i &= \varepsilon_{i} \otimes  \sigma(v_{i+1}) \label{eq:predictionDynamic}\\
        \gamma\,\dot{W}_i &= \sigma(v_{i+1}) \otimes \varepsilon_{i} \label{eq:correctionDynamic}\\
        \gamma\, \dot{b}_i &= \varepsilon_i \label{eq:biasDynamic}
    \end{align}
}

\newcommand{\HNRule}{
\begin{align}
    \vb{W} &= \tfrac{1}{N} \sum\limits_{i=1}^{N} \vb{x}_i \vb{x}_i^T  = \tfrac{1}{N} \vb{X}\vb{X}^{T} \; \text{where } \vb{W}_{ii} = 0 \ , \label{HNWeights} \\
    \vb{b} &= \tfrac{1}{N}\sum\limits_{i=1}^{N} \vb{x}_i \ . \label{HNBias}  
\end{align}
}

\newcommand{\HNsyncUpdate}{
\begin{equation}
    \vb{s}^{t+1} = \mathbf{sgn}(\vb{W} \vb{s}^t +  \vb{b}) \ , \label{HNsyncUpdate}
\end{equation}
}

\newcommand{\HNEnergy}{
\begin{equation}
    E(\vb{s}) = -\tfrac{1}{2} \sum_{j\neq i} s_i \vb{W}_{ij} s_j - \sum_{j=1}^d b_j s_j =  -\left(\tfrac{1}{2} \vb{s}^T \vb{W} \vb{s} + \vb{s}^T \vb{b} \right) \ . \label{HNEnergy}
\end{equation}
}

\newcommand{\HNasyncUpdate}{
\begin{equation}
    s_{i}^{t+1} =
    \begin{cases}
        \phantom{-}1 &\text{if } \sum_{j\neq i} W_{ij} s_j^t\geq 0 \ , \\
        -1 &\text{if } \sum_{j\neq i} W_{ij} s_j^t < 0 \ .
    \end{cases}
    \label{HNasyncUpdate}
\end{equation}
}

\newcommand{\hopfieldEnergyOne}{
\begin{equation}
    E(\vb{s}) = -\tfrac{1}{2} \sum_{j\neq i} s_i W_{ij} s_j - \sum_j b_j s_j \ . \label{HopEnergy}
\end{equation}
}

\newcommand{\hopfieldEnergyTwo}{
\begin{equation}
    E(\vb{s}) = -\tfrac{1}{2} \sum_{i=1}^{N} F(\vb{x}_i \cdot \vb{s})- \sum_{j=1}^d b_j s_j \;. \label{HopEnergy2} 
\end{equation}
}

\newcommand{\hyperparametersTable}{
\begin{table}
  \caption{Hyperparameters Setting used for Experiments}
  \label{hyperparameters}
  \centering
  \begin{tabular}{cc|c|c}
    \toprule
    \multicolumn{2}{c|}{Time Constants} & Error Decay & Time Step  \\
    $\tau$ & $\gamma$  & $\zeta$ & $dt$\\
    \midrule
    $v$: 0.1 & $M$: 0.2 & 1 & 1 ms          \\
    $\varepsilon$: 0.1 & $W$: 0.2 & &  \\
                       & $b$: 0.1 &  & \\
    \bottomrule
  \end{tabular}
\end{table}
}

\newcommand{\clustersProperty}{
\begin{eqnarray}
    &\forall j\neq i,& \mu_{c_i} \perp \mu_{c_j},\ \text{and } \nexists j,\ y_{c_i} = y_{c_j} \; \text{Single-Gaussian} \label{eq:singleGaussian}\\
    &\forall j\neq i,& \mu_{c_i} \perp \mu_{c_j},\ \text{and } \exists j,\ y_{c_i} = y_{c_j} \;  \text{Multi-Gaussian} . \label{eq:multiGaussian}
\end{eqnarray}
}
\newcommand{\figHierarchicalPC}{
\begin{figure}[tbh]
\centering
\includegraphics[width=0.70\columnwidth]{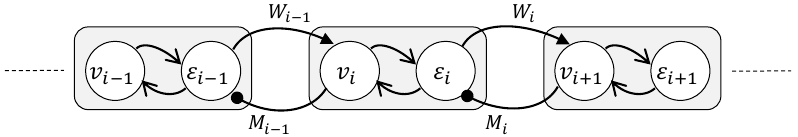}
\caption{A typical PCnet arranged in a feed-forward manner. Each box represents a population of neurons containing value and error nodes.}
\label{fig:PC}
\end{figure}
}

\newcommand{\figAdvExample}{
\begin{figure}[tbh]
    \centering
    \includegraphics[width=0.4\columnwidth]{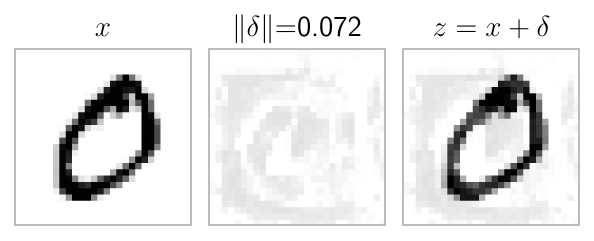}
    \caption{FFnet's perception of the image changes as the noise perturbed the image. FFnet perceives the original image  $\Pr(y_{0}=1|x)=0.99$ while the perception changed to $\Pr(y_{3}=1|x+\delta)=0.87$ on perturbation.}
    \label{fig:advExample}
\end{figure}
}

\newcommand{\figBeforeAfterPC}{
\begin{figure}
    \centering
    \subfloat[\centering Perturbation made to an image.]{\includegraphics[width=.3\columnwidth]{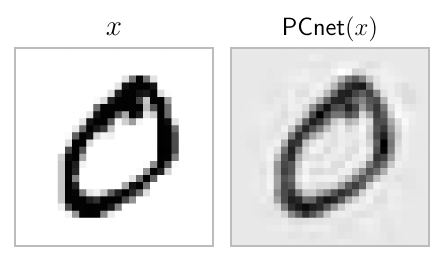}\label{fig:imageBeforeAfterPC}}
    \qquad
    \subfloat[\centering Perturbation made to an adversarial image.]{\includegraphics[width=.3\columnwidth]{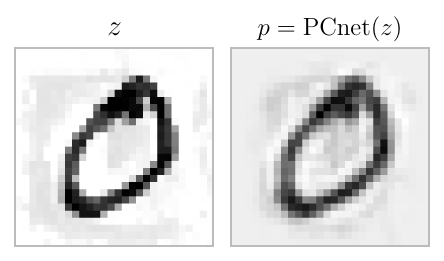}\label{fig:advBeforeAfterPC}}
    \caption{PCnet perturbation is demonstrated using both the original and adversarial images. PCnet modifies the given input based on its trained dynamics. As shown in \protect\subref{fig:imageBeforeAfterPC}, the original image $x$ is depicted on the left, while its perturbation $\mathrm{PCnet}(x)$ is shown on the right. Similarly, \protect\subref{fig:advBeforeAfterPC} presents the adversarial image $z$ on the left, alongside its perturbation $p$ on the right.}
    \label{fig:figBeforeAfterPC}
\end{figure}
}

\newcommand{\figTargetedNonTargetedAdvBeforePC}{
\begin{figure}[tbh]
    \centering
    \includegraphics[width=0.5\columnwidth]{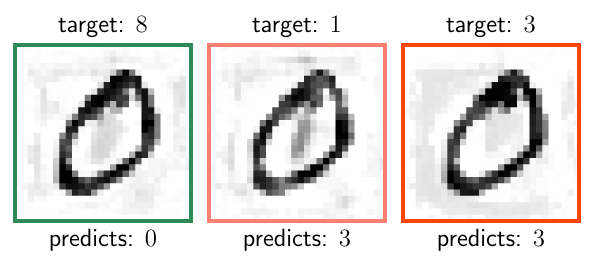}
    \caption{Adversarial attack leads to a successful or failed adversarial example (AE). The attacker may choose a target class so that the FFnet model predicts AE as such, or any attack that causes misclassifications in FFnet is favourable. For the given image of $0$, the attacker aimed for different targets, $8, 1, 3$, from left to right. However, FFnet's prediction might be slightly different from the target. From left to right, the AEs are as follows:
    failed AE (aimed for $8$ and predicted $0$);
    non-targeted AE (aimed for $1$ and predicted $3$);
    targeted AE (aimed for $3$ and predicted $3$).
    Note that even if an AE is non-targeted, it is still a valid AE. Furthermore, the choice of target depends on the attack method. 
    }
    \label{fig:targeted_non_targeted_AdvBeforePC}
\end{figure}
}

\newcommand{\figAdvPCEnergy}{
\begin{figure}[tbh]
    \includegraphics[width=0.80\columnwidth]{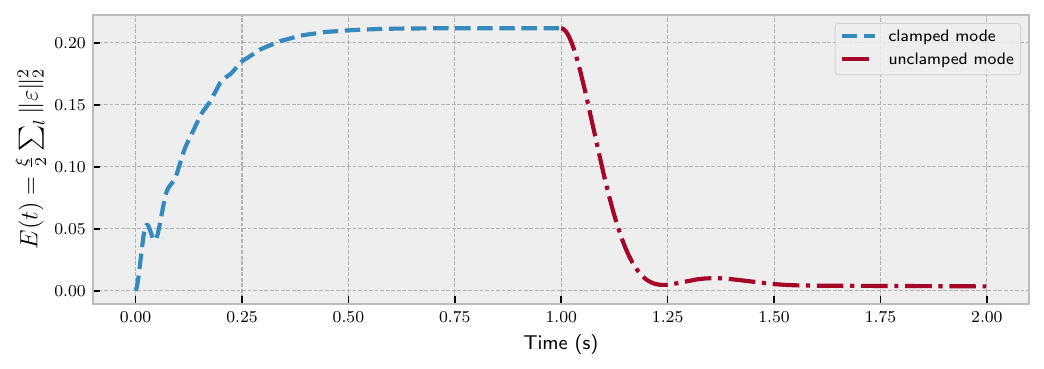}
    \caption{The network's energy drops as PCnet perturbs AE. The network's energy increases and stabilizes when the input is clamped to AE, and it decreases as AE changes through the network's dynamics in PCnet, in unclamp mode.}
    \label{fig:advPCenergy}
\end{figure}
}

\newcommand{\figAdvPredictionCorrectMNIST}{
\begin{figure}[tbh]
    \centering
    \includegraphics[width=.95\columnwidth]{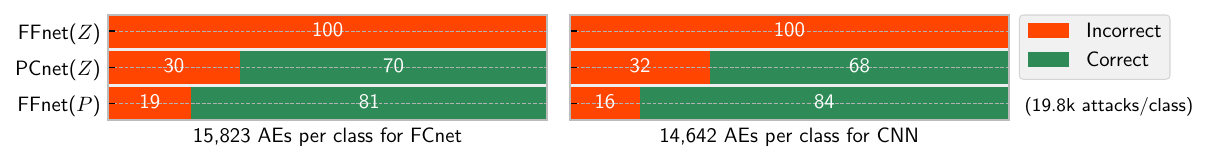}
    \caption{Classification accuracy of AEs on MNIST dataset. For 18k attempted targeted attacks on $X$, using the C\&W method, (1) only AEs that were successful against FFnet were included in these results. (2) PCnet classifies $70\%$ and $68\%$ of AEs for FCnet and CNN, respectively. (3) FFnet classifies $81\%$ and $84\%$ of $P$ (i.e., adjusted AEs by PCnet) for FCnet and CNN, respectively.}
    \label{fig:advPredictionCorrectMNIST}
\end{figure}
}

\newcommand{\figAdvPredictionWrongMNIST}{
\begin{figure}[tbh]
    \centering
    \includegraphics[width=.95\columnwidth]{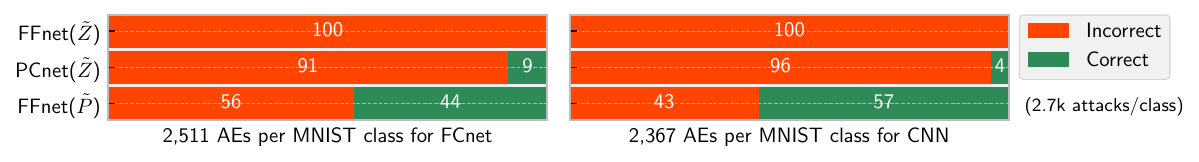}
    \caption{Classification accuracy of AEs on MNIST dataset. For 2.7k attempted targeted attacks on $\tilde{X}$ using the C\&W attack, (1) only AEs that were successful against FFnet were included in these results. (2) PCnet classifies $9\%$ and $4\%$ of AEs for FCnet and CNN, respectively. (3) FFnet classifies $44\%$ and $57\%$ of adjusted $\tilde{Z}$ for FCnet and CNN, respectively.}
    \label{fig:advPredictionWrongMNIST}
\end{figure}
}

\newcommand{\figAdvPredictionCorrectCIFAR}{
\begin{figure}[tbh]
    \centering
    \includegraphics[width=.95\columnwidth]{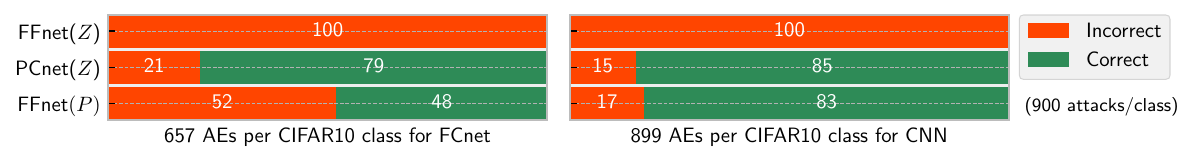}
    \caption{Classification accuracy of AEs on CIFAR10 dataset. For 900 attempted targeted attacks on $X$ using the C\&W attack, (1) only AEs that were successful against FFnet were included in these results. (2) PCnet classifies $79\%$ and $85\%$ of AEs for FCnet and CNN, respectively. (3) FFnet classifies $48\%$ and $83\%$ of $P$ (i.e., adjusted AEs by PCnet) for FCnet and CNN, respectively.}
    \label{fig:advPredictionCorrectCIFAR}
\end{figure}
}

\newcommand{\figAdvPredictionsFFnet}{
\begin{figure}[tbh]
    \centering
\includegraphics[width=.95\columnwidth]{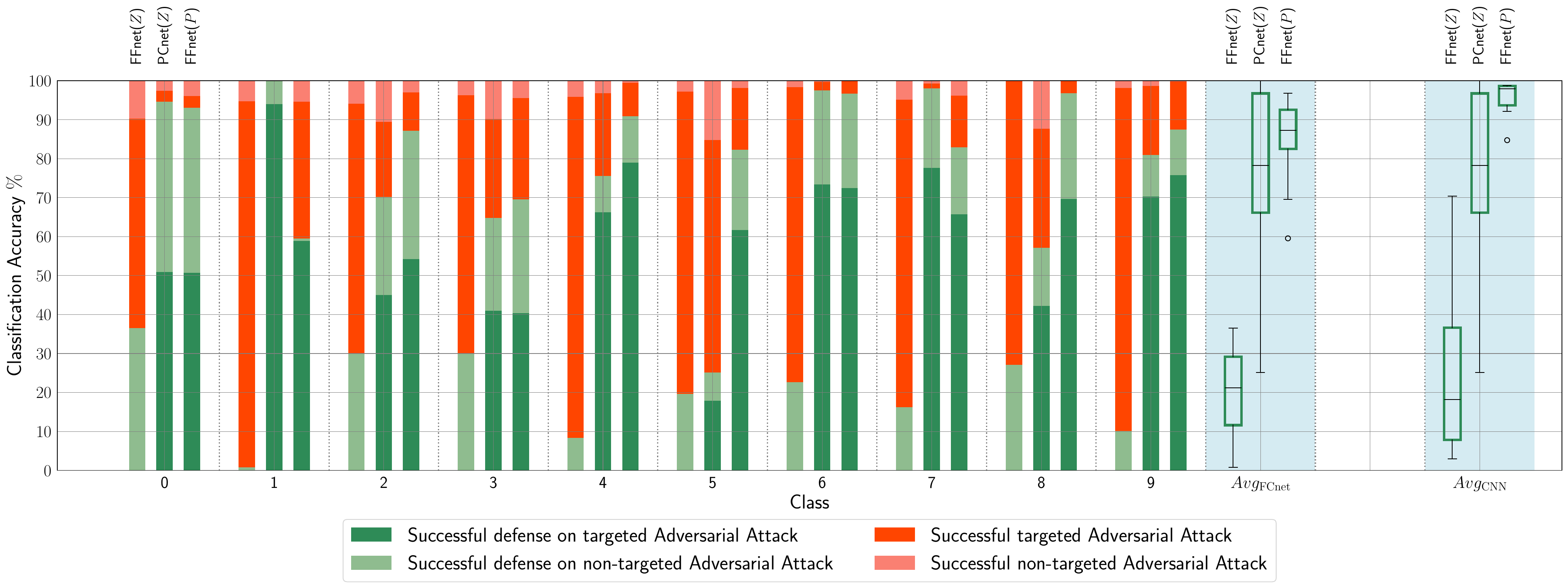}
    \caption{FFnet classifies better on perturbed AEs. In the visual representation, red bars indicate success rates in adversarial attacks, while green bars indicate success rates in classification. For each class, the columns show the classification accuracy of FFnet($Z$), PCnet($Z$), and FFnet($P$), respectively. PCnet performs better than FCnet in classifying targeted and non-targeted adversarial examples (AEs). Additionally, FCnet improves its classification of AEs when PCnet perturbs them. Although the performance of PCnet and FCnet varies across different classes, on average, FCnet performs better on perturbed AEs. Similar behaviour is observed when using CNN. Therefore, FFnets, either FCnet or CCN, show improvement in the classification of AEs when perturbed by PCnet.}
    \label{fig:advPredictionPC&FFnet}
\end{figure}
}

\newcommand{\figAdvPredictionsFC}{
\begin{figure}[tbh]
    \centering
\includegraphics[width=.95\columnwidth]{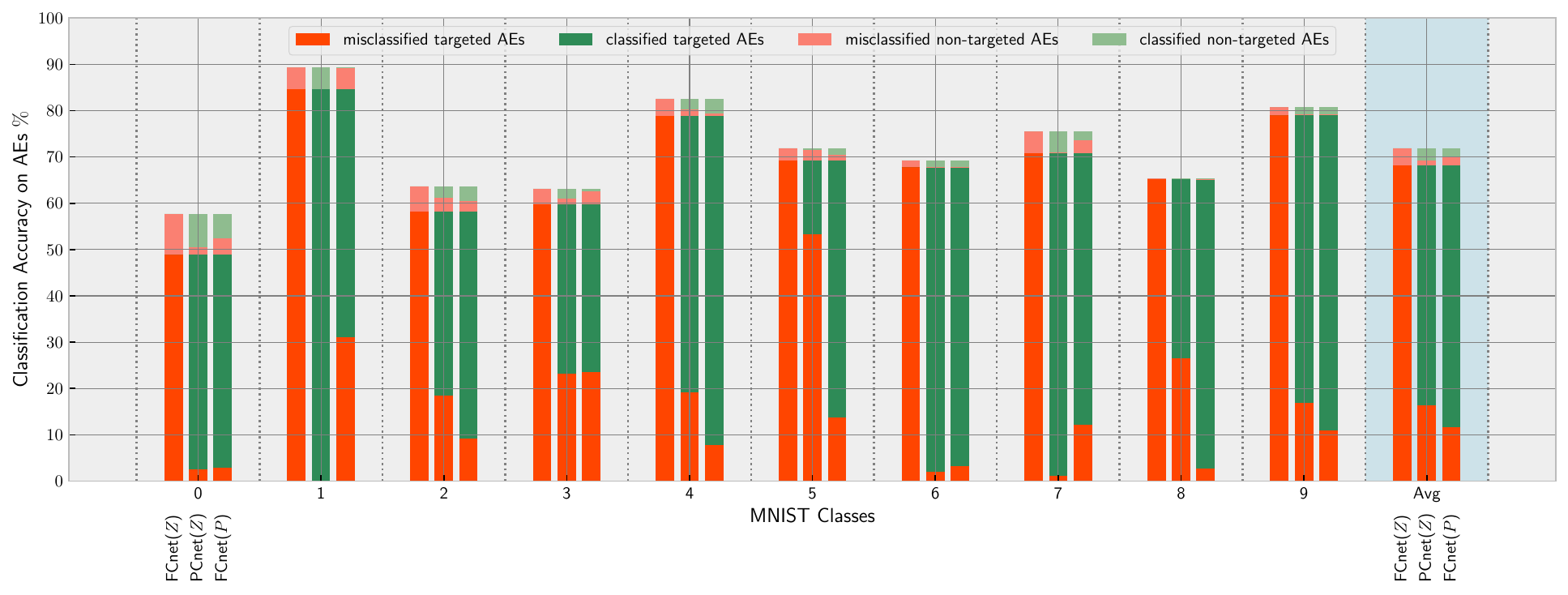}
    \caption{FCnet classifies better on perturbed AEs. In the visual representation, the red bars indicate success rates in adversarial attacks, while the green bars indicate success rates in classification. For each class, the columns show the classification accuracy of FCnet($Z$), PCnet($Z$), and FCnet($P$), respectively. PCnet classifies targeted and non-targeted adversarial examples (AEs) better than FCnet. Additionally, FCnet improves its classification of AEs when PCnet perturbs them. Although the performance of PCnet and FCnet varies across different classes, on average, FCnet performs better on perturbed AEs.}
    \label{fig:advPredictionPC&FC}
\end{figure}
}

\newcommand{\figAdvPredictionsCNN}{
\begin{figure}[tbh]
    \centering
\includegraphics[width=.95\columnwidth]{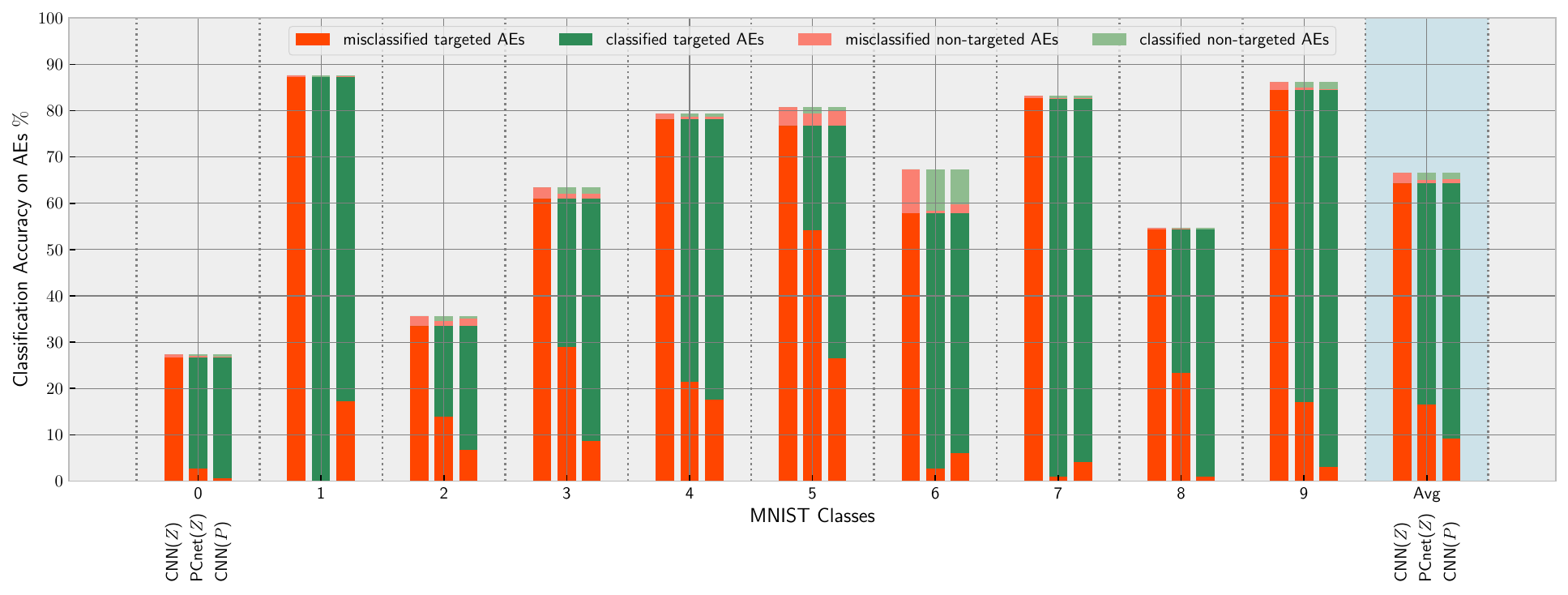}
    \caption{CNN classifies better on perturbed AEs. In the visual representation, the red bars indicate success rates in adversarial attacks, while the green bars indicate success rates in classification. For each class, the columns show the classification accuracy of CNN($Z$), PCnet($Z$), and CNN($P$), respectively. PCnet classifies targeted and non-targeted adversarial examples (AEs) better than CNN. Additionally, CNN improves its classification of AEs when PCnet perturbs them. Although the performance of PCnet and CNN varies across different classes, on average, CNN performs better on perturbed AEs.}
    \label{fig:advPredictionPC&CNN}
\end{figure}
}

\newcommand{\figPCAEsDiagram}{
\begin{figure}[tbh]
    \centering
    \includegraphics[width=.95\columnwidth]{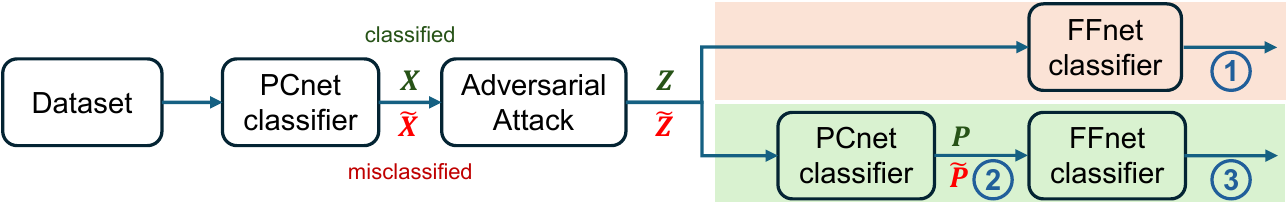}
    \caption{First, the workflow involves dividing the dataset into two parts: $X$ and $\tilde{X}$. After that, we generate Adversarial Examples (AEs) by attacking the FFnet, which can be represented as $\textbf{AT}: X \rightarrow Z$ and $\tilde{X} \rightarrow \tilde{Z}$. We then assess the defence strategy against the attack, first using the AEs directly (i.e., $Z$ and $\tilde{Z}$), and then after making some adjustments to the AEs using the PCnet (i.e., $P$ and $\tilde{P}$).}
    \label{fig:experiment_diagram}
\end{figure}
}

\newcommand{\figPredcitxHat}{
\begin{figure}[tbh]
    \centering
    \includegraphics[width=0.85\columnwidth]{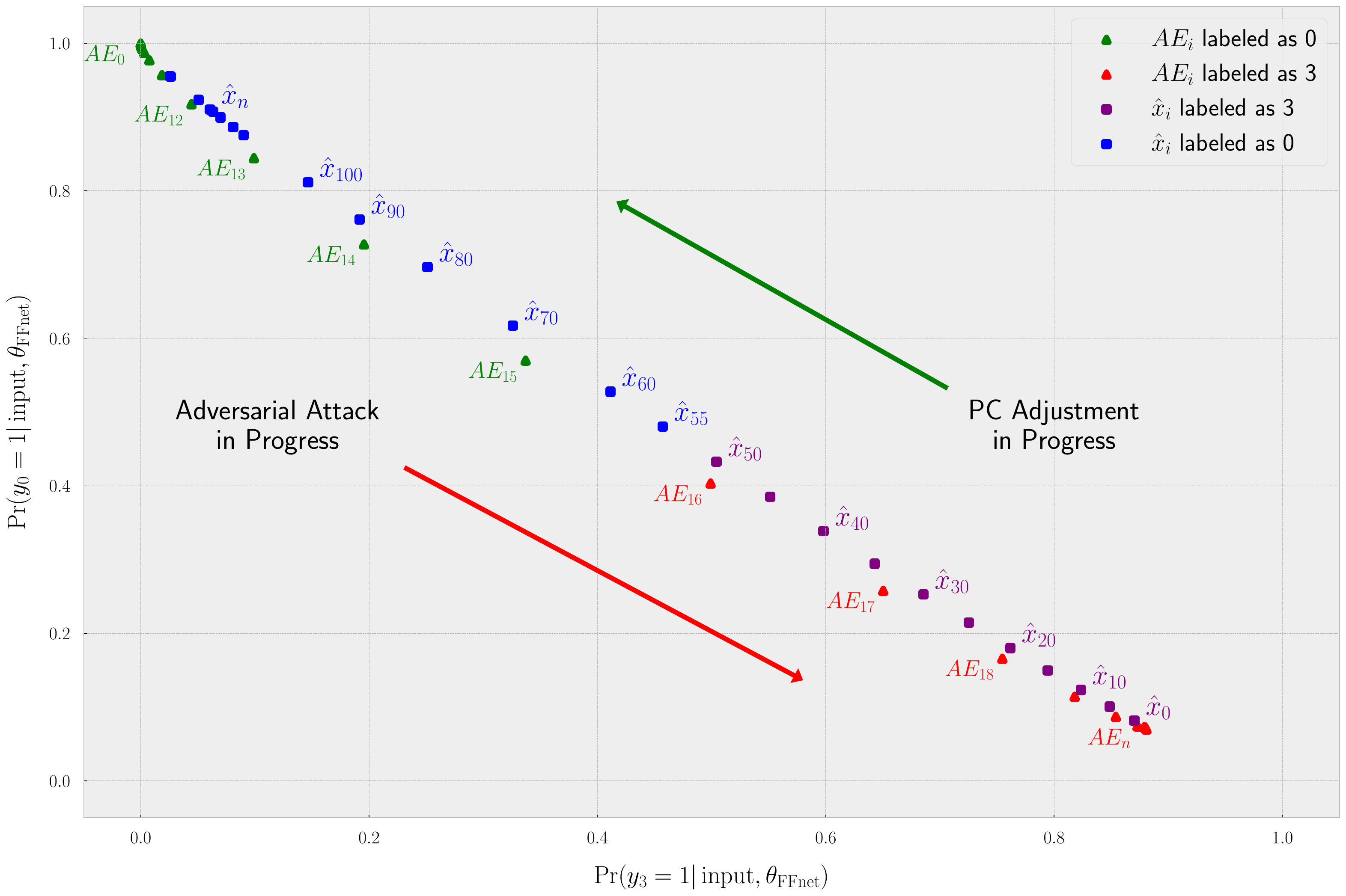}
    \caption{Perturbation and reversion process. The red arrow illustrates FFnet misclassification as perturbed images transition from $AE_0$ to $AE_n$, mistakenly predicting `3' instead of `0'. Conversely, the green arrow depicts PCnet's iterative adjustments, enabling FFnet to correctly classify AEs as `0' after some iterations.}
    \label{fig:FC_adjusted_adv}
\end{figure}
}

\newcommand{\figPerturbRevert}{
\begin{figure}[tbh]
    \centering
    \includegraphics[width=0.70\columnwidth]{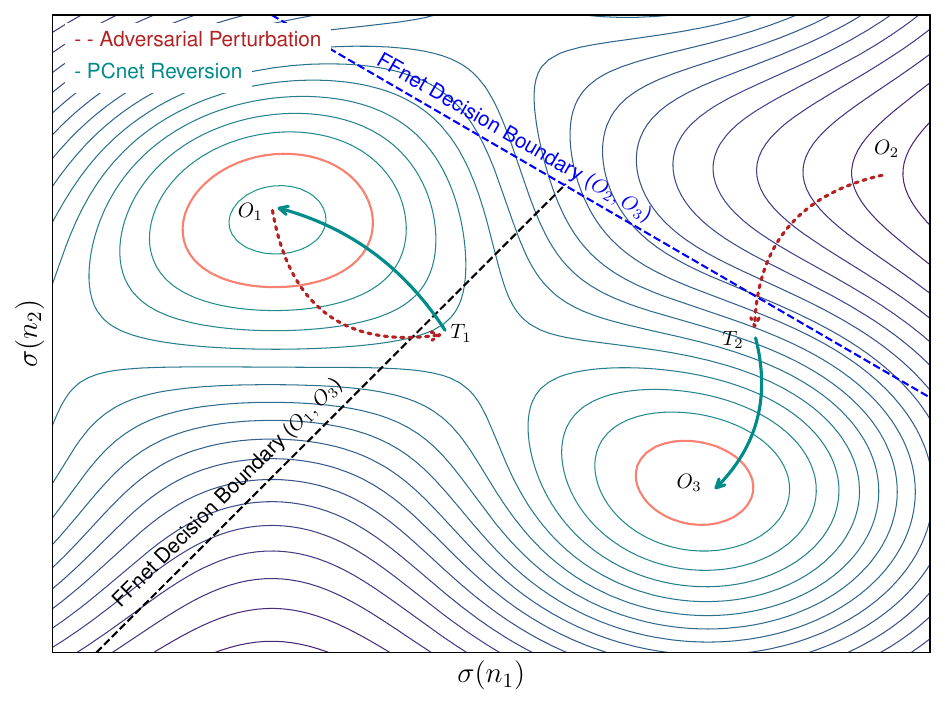}
    \caption{PCnet reverts adversarial attack disguise. The level curves of the network's energy, as a function of the activity of 2 neurons in the network, involve multiple local minima. C\&W attack tries to perturb the original image $O_1$ such that FFnet classifies it as $T_1$. The red curve shows the perturbation path crossing the decision boundary of the FFnet. In the other direction, PCnet, using its dynamics, reverts the perturbation and brings the perturbed image to a state corresponding to lower energy, which is depicted by the green arrow from $T_1$ to $O_1$. When the perturbed image passes the PCnet decision boundary as well, the dynamics push it even further away. For example, instead of returning $T_2$ to $O_2$, it goes toward $O_3$.}
    \label{fig:perturb_reverte}
\end{figure}
}

\newcommand{\figPCvsFilters}{
\begin{figure}[tbh]
    \centering
    \includegraphics[width=1.\columnwidth]{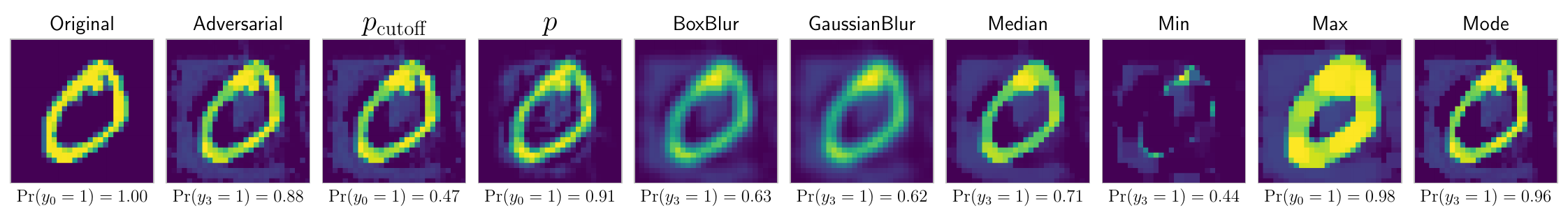}
    \caption{PCnet outperforms standard image filters in reversing adversarial attacks. From left to right, the image sequence consists of the original image, the targeted adversarial image, the minimal perturbed image used to change the prediction, the ultimate perturbation, and six additional filters applied to the adversarial image: Box blur, Gaussian blur, Median, Min, Max, and Mode. The x-axis displays the model's predicted labels and probabilities as $\Pr(y_t=1)$. In these examples, the correct class is $0$ and the targeted label is $3$.}
    \label{fig:PCnet_vs_filters}
\end{figure}
}

\newcommand{\figPCvsFiltersHot}{
\begin{figure}[tbh]
    \centering
    \includegraphics[width=0.80\columnwidth]{figs/filters_hot_0_3.pdf}
    \caption{TBA.}
    \label{fig:PCnet_vs_filters_hot}
\end{figure}
}

\newcommand{\figPCvsFiltersCosine}{
\begin{figure}[tbh]
    \centering
    \includegraphics[width=0.5\columnwidth]{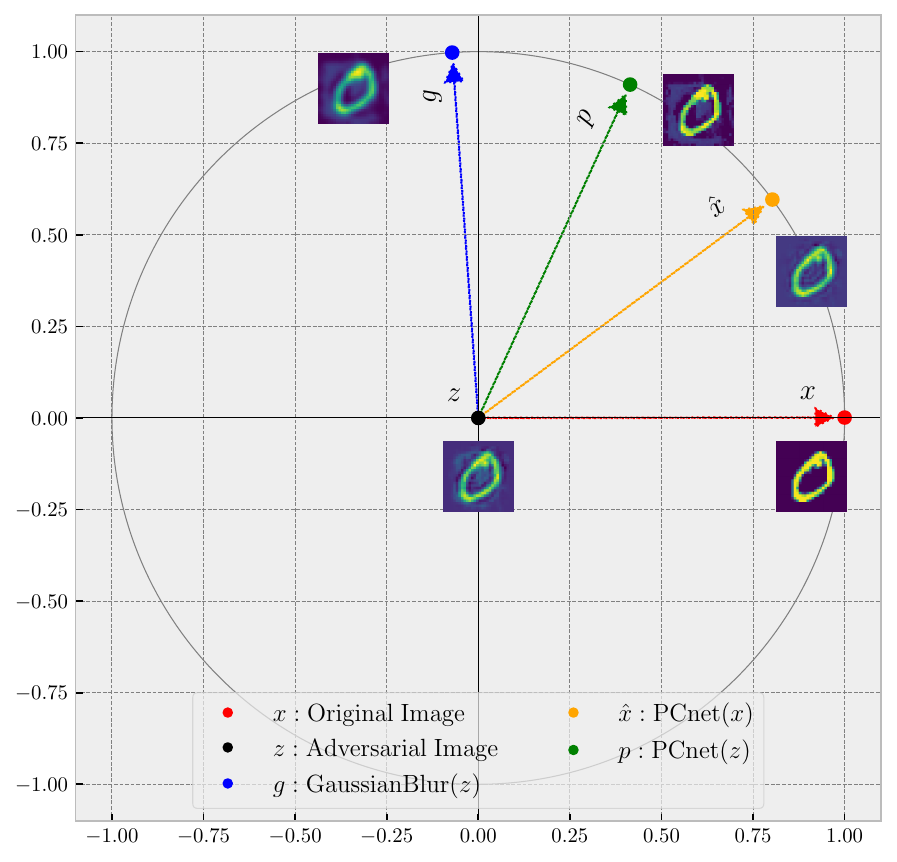}
    \caption{The modified AE closely resembles the original. Mapping directions from targeted AE, $z$, to the original image, $x$, to Gaussian-filtered one, $g=G(z)$, and to perturbed AE by PCnet, $p$, presented in red, blue and green vectors. The perturbed AE by PCnet is more similar to the original image than the Gaussian-filtered one.}
    \label{fig:PCnet_vs_filters_cosine}
\end{figure}
}

\newcommand{\figPCvsFiltersCorrectFCBoxplotMNISTCIFAR}{
\begin{figure}[tbh]
    \centering
    \includegraphics[width=1.0\columnwidth]{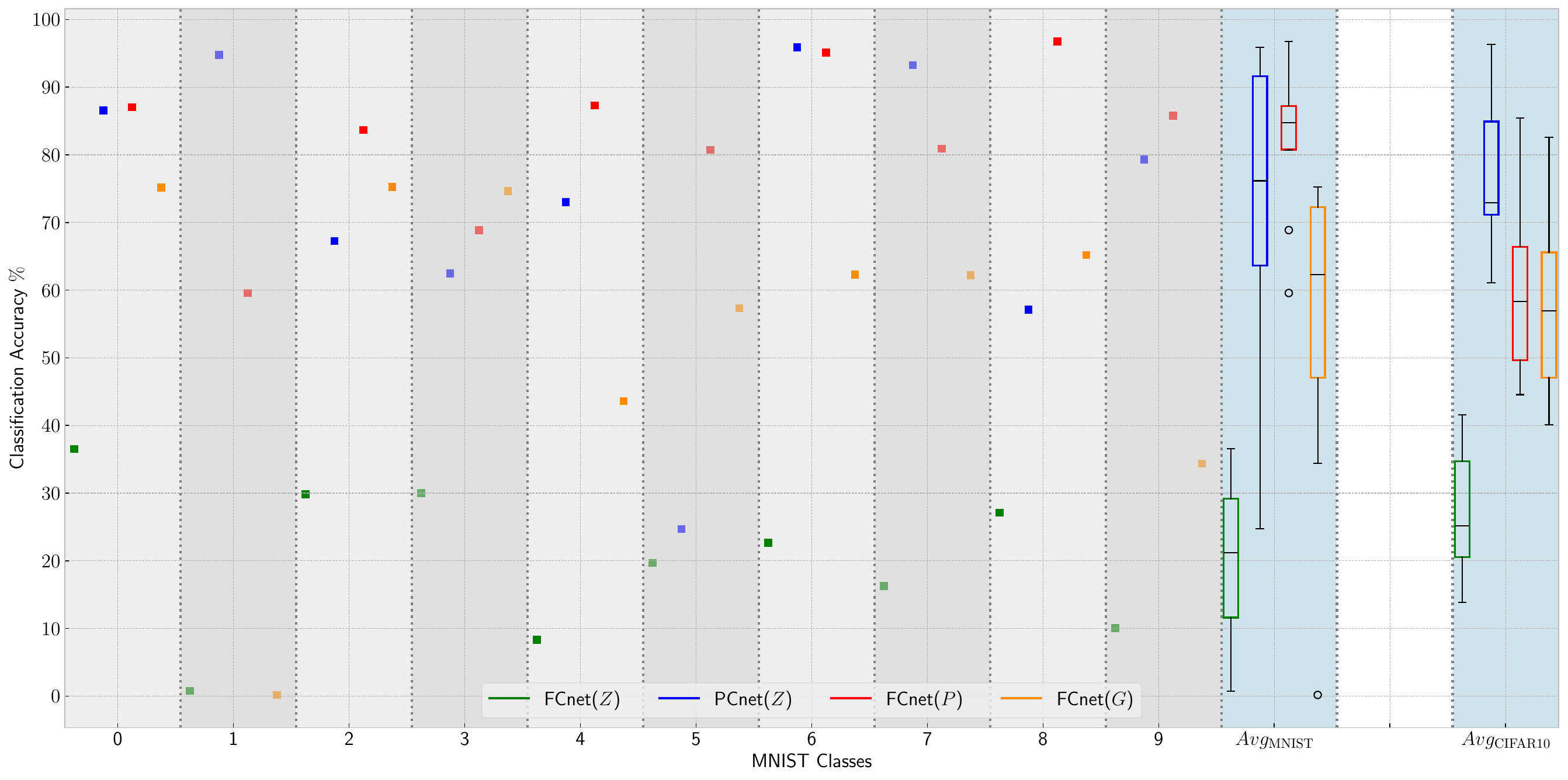}
    \caption{PCnet's perturbation outperforms filter effect on classifying AEs. When AEs, created using classified data by the PCnet, are processed with a Gaussian filter, the classification accuracy of the FCnet improves. However, in most cases, the perturbation caused by the PCnet has a greater impact on FCnet classification than the effects of the Gaussian filter. On average, FCnet performs better on AEs perturbed by PCnet than when a Gaussian filter is used. This trend is consistent for AEs from both the MNIST and CIFAR10 datasets.
    Note that:
    $Z\leftarrow\mathrm{AT_{FCnet}}(X)$,\;
    $G\leftarrow\mathrm{GaussianBlur}(Z)$,\;
    $\left(P, \textrm{and labels}\right) \leftarrow \mathrm{PCnet}(Z)$,\;
    $\textrm{labels} \leftarrow \mathrm{FCnet}(\cdot)$.
    }
    \label{fig:PCnet_filter_boxplot_MNIST_CIFAR10_FC_correct}
\end{figure}
}

\newcommand{\figPCvsFiltersCorrectCNNBoxplotMNISTCIFAR}{
\begin{figure}[tbh]
    \centering
    \includegraphics[width=1.0\columnwidth]{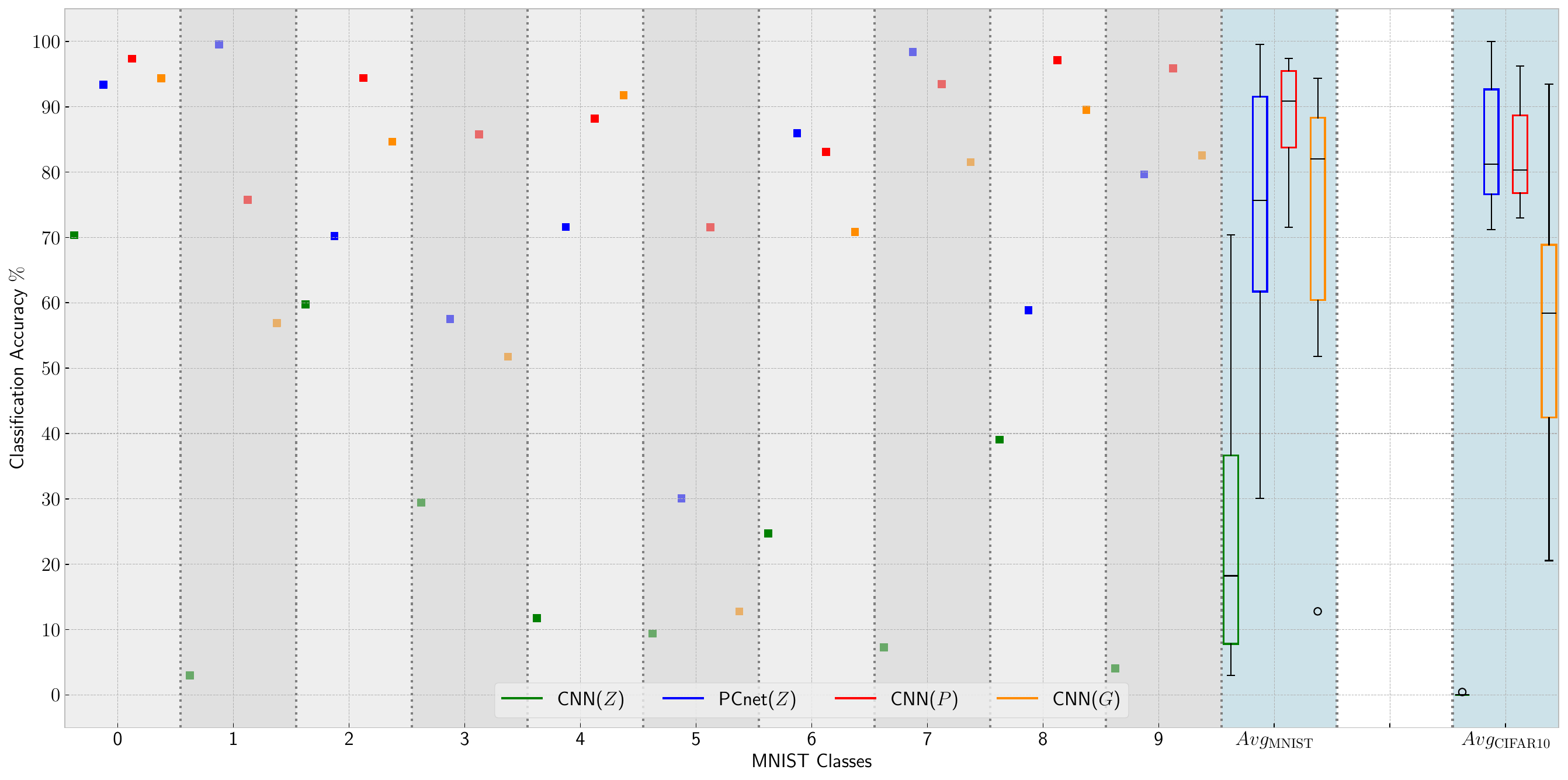}
    \caption{PCnet's perturbation outperforms filter effect on classifying AEs. When AEs, created using classified data by the PCnet, are processed with a Gaussian filter, the classification accuracy of the CNN improves. However, in most cases, the perturbation caused by the PCnet has a greater impact on CNN classification than the effects of the Gaussian filter. On average, CNN performs better on AEs perturbed by PCnet than when a Gaussian filter is used. This trend is consistent for AEs from both the MNIST and CIFAR10 datasets.
    Note that:
    $Z\leftarrow\mathrm{AT_{CNN}}(X)$,\;
    $G\leftarrow\mathrm{GaussianBlur}(Z)$,\;
    $\left(P, \textrm{and labels}\right) \leftarrow \mathrm{PCnet}(Z)$,\;
    $\textrm{labels} \leftarrow \mathrm{CNN}(\cdot)$.
    }
    \label{fig:PCnet_filter_boxplot_MNIST_CIFAR10_CNN_correct}
\end{figure}
}

\newcommand{\figPCvsFiltersWrongFCBoxplotMNISTCIFAR}{
\begin{figure}[tbh]
    \centering
    \includegraphics[width=1.0\columnwidth]{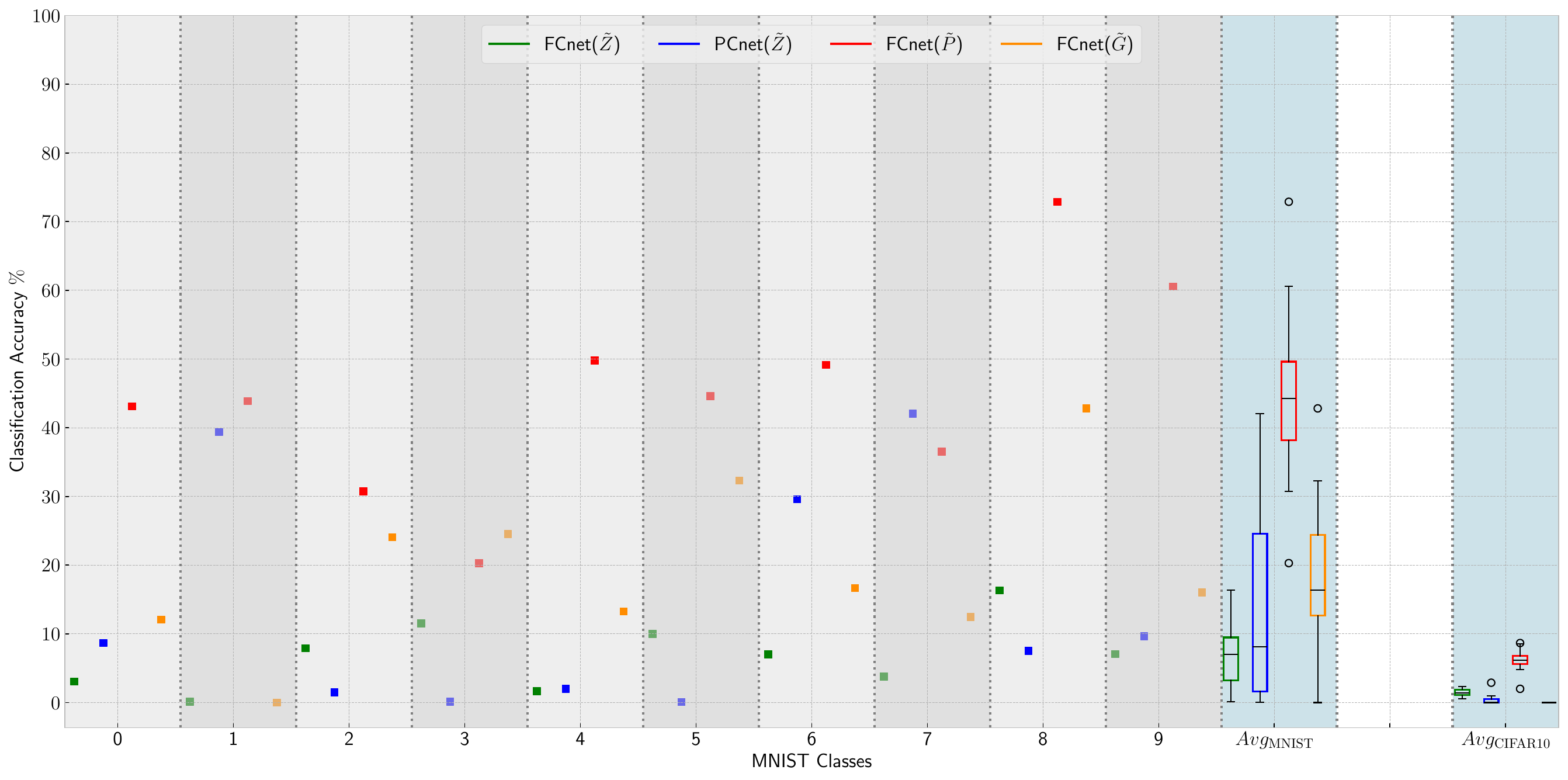}
    \caption{PCnet's perturbation outperforms filter effect on classifying AEs. When AEs, created using misclassified data by the PCnet, are processed with a Gaussian filter, the classification accuracy of the FCnet improves. However, in most cases, the perturbation caused by the PCnet has a greater impact on FCnet classification than the effects of the Gaussian filter. On average, FCnet performs better on AEs perturbed by PCnet than when a Gaussian filter is used. This trend is consistent for AEs from both the MNIST and CIFAR10 datasets.
    Note that:
    $\tilde{Z}\leftarrow\mathrm{AT_{FCnet}}(\tilde{X})$,\;
    $\tilde{G}\leftarrow\mathrm{GaussianBlur}(\tilde{Z})$,\;
    $\left(\tilde{P}, \textrm{and labels}\right) \leftarrow \mathrm{PCnet}(\tilde{Z})$,\;
    $\textrm{labels} \leftarrow \mathrm{FCnet}(\cdot)$.
    }
    \label{fig:PCnet_filter_boxplot_MNIST_CIFAR10_FC_wrong}
\end{figure}
}

\newcommand{\figPCvsFiltersWrongCNNBoxplotMNISTCIFAR}{
\begin{figure}[tbh]
    \centering
    \includegraphics[width=1.0\columnwidth]{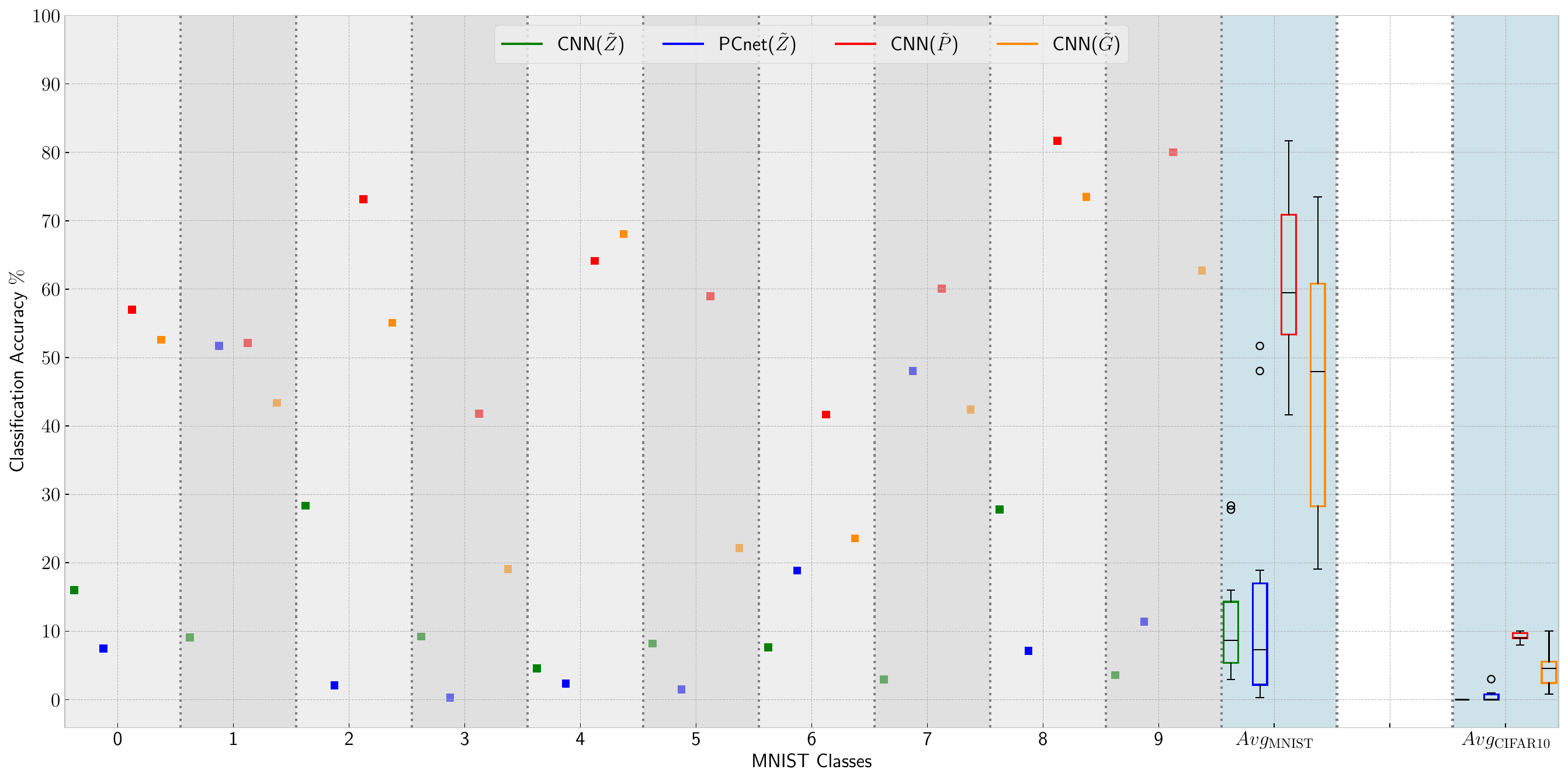}
    \caption{PCnet's perturbation outperforms filter effect on classifying AEs. When AEs, created using misclassified data by the PCnet, are processed with a Gaussian filter, the classification accuracy of the CNN improves. However, in most cases, the perturbation caused by the PCnet has a greater impact on CNN classification than the effects of the Gaussian filter. On average, CNN performs better on AEs perturbed by PCnet than when a Gaussian filter is used. This trend is consistent for AEs from both the MNIST and CIFAR10 datasets. 
    Note that:
    $\tilde{Z}\leftarrow\mathrm{AT_{CNN}}(\tilde{X})$,\;
    $\tilde{G}\leftarrow\mathrm{GaussianBlur}(\tilde{Z})$,\;
    $\left(\tilde{P}, \textrm{and labels}\right) \leftarrow \mathrm{PCnet}(\tilde{Z})$,\;
    $\textrm{labels} \leftarrow \mathrm{CNN}(\cdot)$.
    }
    \label{fig:PCnet_filter_boxplot_MNIST_CIFAR10_CNN_wrong}
\end{figure}
}

\newcommand{\figSankeyFCnetClassifiedMNIST}{
\begin{figure}[tbh]
    \centering
    \includegraphics[width=0.99\columnwidth]{figs/sankey_adversaarial_classified_mnist_FCnet.pdf}
    \caption{Classification improves over AEs. For the given classified images of the MNIST dataset, the classification improves through the following phases: (1) creating AEs using C\&W Adversarial attacks on FCnet; (2) classifying AEs using PCnet; (3) classifying perturbed AEs using FCnet.}
    \label{fig:AE_FCnet_classifiedMNIST}
\end{figure}
}

\newcommand{\figSankeyCNNClassifiedMNIST}{
\begin{figure}[tbh]
    \centering
    \includegraphics[width=0.99\columnwidth]{figs/sankey_adversaarial_classified_mnist_CNN.pdf}
    \caption{Classification improves over AEs. For the given classified images of the MNIST dataset, the classification improves through the following phases: (1) creating AEs using C\&W Adversarial attacks on CNN; (2) classifying AEs using PCnet; (3) classifying perturbed AEs using CNN.}
    \label{fig:AE_CNN_classifiedMNIST}
\end{figure}
}

\newcommand{\figSankeyFCnetmisClassifiedMNIST}{
\begin{figure}[tbh]
    \centering
    \includegraphics[width=0.99\columnwidth]{figs/sankey_adversaarial_misclassified_mnist_FCnet.pdf}
    \caption{Classification improves over AEs. For the given misclassified images of the MNIST dataset, the classification improves through the following phases: (1) creating AEs using C\&W Adversarial attacks on FCnet; (2) classifying AEs using PCnet; (3) classifying perturbed AEs using FCnet.}
    \label{fig:AE_FCnet_misclassifiedMNIST}
\end{figure}
}

\newcommand{\figSankeyCNNmisClassifiedMNIST}{
\begin{figure}[tbh]
    \centering
    \includegraphics[width=0.99\columnwidth]{figs/sankey_adversaarial_misclassified_mnist_CNN.pdf}
    \caption{Classification improves over AEs. For the given misclassified images of the MNIST dataset, the classification improves through the following phases: (1) creating AEs using C\&W Adversarial attacks on CNN; (2) classifying AEs using PCnet; (3) classifying perturbed AEs using CNN.}
    \label{fig:AE_CNN_misclassifiedMNIST}
\end{figure}
}

\newcommand{\figAdvLoss}{
\begin{figure}[tb]
    \centering
    \includegraphics[width=\columnwidth]{figs/fc_adv_loss.pdf}
    \caption{Part of a traditional PC network. Each circle represents a population of neurons.}
    \label{fig:advLoss}
\end{figure}
}

\newcommand{\figAdvProb}{
\begin{figure}[th]
    \centering
    \includegraphics[width=\columnwidth]{figs/fc_adv_prob.pdf}
    \caption{Part of a traditional PC network. Each circle represents a population of neurons.}
    \label{fig:advLoss}
\end{figure}
}

\newcommand{\figModelsPerformance}{
\begin{figure}[tb]
    \centering
    \includegraphics[width=\columnwidth]{figs/ComparisonChart.pdf}
    \caption{Part of a traditional PC network. Each circle represents a population of neurons.}
    \label{fig:None}
\end{figure}
}

\newcommand{\figTreemap}{
\begin{figure}
\centering
\begin{tikzpicture}
    \begin{axis}[
        xbar stacked, 
        xmin=0, xmax=100, 
        xtick={0,20,40,60,80,100},
        ytick=data, 
        legend style={at={(0.5,1.05)}, anchor=south, legend columns=-1}, 
        legend cell align=left, 
        width=\columnwidth,
        height=4cm,
        bar width=5mm, 
        enlarge y limits=0.5, 
        symbolic y coords={Phase 1, Phase 2, Phase 3}, 
        nodes near coords, 
        nodes near coords align={center}, 
    ]
        \addplot coordinates {(68.29,Phase 1) (17.3,Phase 2) (11.75,Phase 3)};
        \addlegendentry{A}
    
        \addplot coordinates {(10,Phase 1) (60.99,Phase 2) (66.54,Phase 3)};
        \addlegendentry{B}
    
        \addplot coordinates {(3.6,Phase 1) (5.04,Phase 2) (2.08,Phase 3)};
        \addlegendentry{C}
    
        \addplot coordinates {(18.11,Phase 1) (16.67,Phase 2) (19.63,Phase 3)};
        \addlegendentry{D}
    \end{axis}
\end{tikzpicture}
\caption{Horizontal distribution of Parts \color{blue}A\color{black}, \color{red}B\color{black}, \color{brown}{C}\color{black}, and \color{darkgray}D\color{black} across different models in percentages.}
\end{figure}
}

\newcommand{\tabABCD}{
\begin{table}
    \caption{Every adversarial example can be categorized into one of four categories based on the classifier's performance and the attacker's intention.}
    \label{tab:ABCD}
    \centering


    

    \begin{tabular}{r|@{}c@{}c@{}}
        & \multicolumn{2}{c}{\textbf{Adversarial Attack}} \\
        & \makebox[5em]{targeted}
        & \makebox[5em]{non-targeted} \\ \hline
    classified
        & \colorbox{colorB}{\makebox[4em][c]{\textcolor{white}{B}}}
        & \colorbox{colorD}{\makebox[4em][c]{\textcolor{white}{D}}} \\
    misclassified
        & \colorbox{colorA}{\makebox[4em][c]{\textcolor{white}{A}}}
        & \colorbox{colorC}{\makebox[4em][c]{\textcolor{white}{C}}}
    \end{tabular}

\end{table}
}

\newcommand{\tabComputationalCost}{
\begin{table}
    \caption{required time to create each adversarial example using different methods.}
    \label{tab:computationalCost}
    \centering
    \begin{tabular}{r|lll}
    Attacks & \multicolumn{2}{c}{{\bf Computational Cost (per adversarial example)}} \\ \hline
       FGSM & $3.12 ms \pm  104 \mu s$ per loop & ($\mu \pm \sigma$ of 7 runs, 10 loops each) \\
        BIM & $28.0 ms \pm 4.68 ms$ per loop & ($\mu \pm \sigma$ of 7 runs, 10 loops each) \\ 
        PGD & $90.5 ms \pm 2.16 ms$ per loop & ($\mu \pm \sigma$ of 7 runs, 10 loops each) \\
        C\&W & $452 ms \pm 10.9 ms$ per loop & ($\mu \pm \sigma$ of 7 runs, one loop each)
    \end{tabular}
\end{table}
}

\newcommand{\tabTreemap}{
\begin{table}[!ht]
    \centering
    \caption{This is caption}
    \begin{tabular}{l|rrrr}
    \hline
    \textbf{AEs Distribution} & \multicolumn{1}{c}{\color{blue}\textbf{A}}\color{black} & \multicolumn{1}{c}{\color{red}\textbf{B}}\color{black} & \multicolumn{1}{c}{\color{brown}\textbf{C}}\color{black} & \multicolumn{1}{c}{\color{darkgray}\textbf{D}}\color{black} \\
    \hline
        \textbf{FCnet on AEs} & 68.29\% & 10.00\% & 3.60\% & 18.11\% \\ 
        \textbf{PCnet on AEs} & 17.30\% & 60.99\% & 5.04\% & 16.67\% \\ 
        \textbf{FFnet on Adj. AEs} & 11.75\% & 66.54\% & 2.08\% & 19.93\%
    \end{tabular}
\end{table}
}

\newcommand{\tabPCnetPercisionMNIST}{
\begin{table}
    \caption{PCnet Precision on Super Stimulus MNIST digits using FFnet.}
    \label{tab:pcClassifier}
    \centering\hrule
    \resizebox{\columnwidth}{!}{%
    \begin{tabular}{rcc}
        \toprule
        PCnet/data & regular data & superstimulus data\\
        Precision &  $76.82\%$ & $83.08\%$
        \bottomrule
    \end{tabular}
    }\hrule
\end{table}
}

\newcommand{\tabMNISTModels}{
\begin{table}[ht]
  \caption{Models' Hyper-parameters for experiments on MNIST dataset}
  \label{tab:MNISTmodels}
  \centering
  \begin{tabular}{cccc|cc}
    \toprule
    Model & Layers & Batch size, \#epochs, lr & \# train/test & Accuracy (\%) \\
    \midrule
    PCnet & 784-128-64-32-16-10 & 64, 5, 10^{-3} & 6k, 10k & $74.28, 74.22$ \\
    FCnet & 784-50-20-10 & 64, 5, 10^{-3} & 60k, 10k & $92.99, 92.82$ \\
    CNN & conv1-pool-conv2-fc1-fc2 & 64, 5, 10^{-3} & 60k, 10k & $96.27, 96.50$ \\
    \midrule
    Details & \multicolumn{4}{|l}{conv1, conv2: Conv2d(kernel size: $5$, stride: $1$, padding: $2$)} \\
    on CNN & \multicolumn{4}{|l}{pool: MaxPool2d(kernel size: $2$, stride: $2$, padding: $0$), fc1($490, 32$), fc2($32, 10$)} \\
    \bottomrule
  \end{tabular}
\end{table}
}

\newcommand{\tabCIFARModels}{
\begin{table}[ht]
  \caption{Models' Hyper-parameters for experiments on CIFAR10 dataset}
  \label{tab:CIFARmodels}
  \centering
  \begin{tabular}{cccc|c}
    \toprule
    Model & Layers & Batch size, \#epochs, lr & \# train/test & Accuracy (\%) \\
    \midrule
    PCnet & $3072-512-16-10$ & $64, 5, 10^{-3}$ & $5k, 10k$ & $22.08, 22.59$ \\
    FCnet & $3072-512-10$ & $128, 10, 10^{-3}$ & $50k, 10k$ & $70.79, 52.04$ \\
    CNN & conv1-conv2-pool-fc1-fc2 & $128, 10, 10^{-3}$ & $50k, 10k$ & $98.48, 72.52$ \\
    \midrule
    Details & \multicolumn{4}{|l}{conv1, conv2: Conv2d(kernel size: $3$, stride: $1$, padding: $1$)} \\
    on CNN & \multicolumn{4}{|l}{pool: MaxPool2d(kernel size: $2$, stride: $2$, padding: $0$), fc1($4096, 512$), fc2($512, 10$)} \\
    \bottomrule
  \end{tabular}
\end{table}
}

\newcommand{\tabModelsOne}{
\begin{table}[ht]
  \caption{Models' Hyper-parameters for experiments on MNIST dataset}
  \label{tab:MNISTmodels}
  \centering
  \begin{tabular}{cccc|cc}
    \toprule
    Model & Layers & Batch size, \#epochs, lr & \# train/test & Accuracy (\%) \\
    \midrule
    PCnet & $784-128-64-32-16-10$ & $64, 5, 10^{-3}$ & $6k, 10k$ & $74.28, 74.22$ \\
    FCnet & $784-50-20-10$ & $64, 5, 10^{-3}$ & $60k, 10k$ & $92.99, 92.82$ \\
    CNN & conv1-pool-conv2-fc1-fc2 & $64, 5, 10^{-3}$ & $60k, 10k$ & $96.27, 96.50$ \\
    \midrule
    Details & \multicolumn{4}{|l}{conv1, conv2: Conv2d(kernel size: $5$, stride: $1$, padding: $2$)} \\
    on CNN & \multicolumn{4}{|l}{pool: MaxPool2d(kernel size: $2$, stride: $2$, padding: $0$), fc1($490, 32$), fc2($32, 10$)} \\
    \midrule \midrule
    \multicolumn{5}{c}{Models' Hyper-parameters for experiments on CIFAR10 dataset} \\
    \midrule
    PCnet & 3072-512-16-10 & 64, 5, 10^{-3} & 5k, 10k & $22.08, 22.59$ \\
    FCnet & 3072-512-10 & 128, 10, 10^{-3} & 50k, 10k & $70.79, 52.04$ \\
    CNN & conv1-conv2-pool-fc1-fc2 & 128, 10, 10^{-3} & 50k, 10k & $98.48, 72.52$ \\
    \midrule
    Details & \multicolumn{4}{|l}{conv1, conv2: Conv2d(kernel size: $3$, stride: $1$, padding: $1$)} \\
    on CNN & \multicolumn{4}{|l}{pool: MaxPool2d(kernel size: $2$, stride: $2$, padding: $0$), fc1($4096, 512$), fc2($512, 10$)} \\
    \bottomrule
  \end{tabular}
\end{table}
}

\newcommand{\tabModelsTwo}{
\begin{table}[ht]
  \caption{Models' Hyper-parameters for experiments on MNIST and CIFAR10 datasets}
  \label{tab:MNISTmodels}
  \centering
  \begin{tabular}{c|ccc|cc}
    \toprule
     & Model & Layers & Batch size, \#epochs, lr & \# train/test & Accuracy (\%) \\
    \hline
    \multirow{6}{*}{\rotatebox[origin=c]{90}{\parbox[c]{.5cm}{\centering MNIST}}}
    & PCnet & $784-128-64-32-16-10$ & $64, 5, 10^{-3}$ & $6k, 10k$ & $74.28, 74.22$ \\
    & FCnet & $784-50-20-10$ & $64, 5, 10^{-3}$ & $60k, 10k$ & $92.99, 92.82$ \\
    & CNN & conv1-pool-conv2-fc1-fc2 & $64, 5, 10^{-3}$ & $60k, 10k$ & $96.27, 96.50$ \\
    \cline{2-6} 
    & CNN & \multicolumn{4}{|l}{conv1, conv2: Conv2d(kernel size: $5$, stride: $1$, padding: $2$)} \\
    & Details & \multicolumn{4}{|l}{pool: MaxPool2d(kernel size: $2$, stride: $2$, padding: $0$), fc1($490, 32$), fc2($32, 10$)} \\
    \midrule \multicolumn{6}{c}{}\\ \midrule
    \multirow{6}{*}{\rotatebox[origin=c]{90}{\parbox[c]{1cm}{\centering CIFAR10}}}
    & PCnet & $3072-512-16-10$ & $64, 5, 10^{-3}$ & $5k, 10k$ & $22.08, 22.59$ \\
    & FCnet & $3072-512-10$ & $128, 10, 10^{-3}$ & $50k, 10k$ & $70.79, 52.04$ \\
    & CNN & conv1-conv2-pool-fc1-fc2 & $128, 10, 10^{-3}$ & $50k, 10k$ & $98.48, 72.52$ \\
    \cline{2-6}
    & CNN & \multicolumn{4}{|l}{conv1, conv2: Conv2d(kernel size: $3$, stride: $1$, padding: $1$)} \\
    & Details & \multicolumn{4}{|l}{pool: MaxPool2d(kernel size: $2$, stride: $2$, padding: $0$), fc1($4096, 512$), fc2($512, 10$)} \\
    \bottomrule
  \end{tabular}
\end{table}
}

\newcommand{\tabCWmissClass}{
\begin{table}[!ht]
    \caption{Models mistakes in three phases on targeted/non-targeted AEs using $X$ and $\tilde{X}$ for FFnets.}
    \label{tab:CWonAEs}
    \centering
    \begin{tabular}{l|rrr|rrr|r}
    & \multicolumn{3}{c|}{\textbf{Successful Attacks on FCnet}}
    & \multicolumn{3}{c|}{\textbf{Successful Attacks on CNN}} & \\
     failure (\%) &  Phase 1 & Phase 2 & Phase 3 & Phase 1 & Phase 2 & Phase 3 & \# Attacks \\
    \hline
        AE on $X$ & 79.88\% & 23.61\% & 15.33\% & 74.12\% & 23.61\% & 4.30\% & 18k/digit\\ 
        AE on $\tilde{X}$ & 93.20\% & 85.09\% & 52.79\% & 88.27\% & 84.32\% & 36.53\% & 2.7k/digit\\
    \bottomrule
    \end{tabular}
\end{table}
}

\newcommand{\tabSuccessfullAttacks}{
\begin{table}[!ht]
    \caption{On the MNIST and CIFAR10 datasets, models' misclassifications vary on AEs corresponding to $X$ and $\tilde{X}$ for FFnets throughout the three phases. The adversarial attack uses C\&W, FGSM, PGD, and BIM algorithms (with $\epsilon = 0.75$). `Others' represents the average of the last three. Not all AEs, denoted as $Z$, are effective against the adversary. For instance, in the first row and second column, FCnet misclassified $79.91\%$ of the attacks. The other columns display the misclassification rates of $ \textrm{PCnet}(Z)$ and $\textrm{FCnet}(P)$, where $P$ represents the perturbed AEs by PCnet. Likewise, The last three columns present the same metric using the CNN model.}
    \label{tab:SuccessfullAttacks}
    \centering    
    \begin{tabular}{l|rrr|rrr|r|cl}
    & \multicolumn{3}{c|}{\textbf{Successful Attacks on FCnet}} & \multicolumn{3}{c|}{\textbf{Successful Attacks on CNN}} & \\
    Attack &  $\textrm{FC}(Z)$ & $\textrm{PCnet}(Z)$ & $\textrm{FC}(P)$ & $\textrm{CNN}(Z)$ & $\textrm{PCnet}(Z)$ & $\textrm{CNN}(P)$ & \# Attacks & \multicolumn{2}{c}{Data} \\
    \hline \hline
        \myg C\&W & \myo $79.91$\% & $23.79$\% & \myl $15.42$\% & \myo $73.95$\% & $23.86$\% & \myl $11.60$\% & \myg $19.8k/\textrm{class}$ & \multirow{2}{*}{$X$} & \multirow{4}{*}{\rotatebox[origin=c]{90}{\parbox[c]{1cm}{\centering MNIST}}} \\
        others  & \myo $29.88$\% & $11.77$\% & \myl $12.25$\% & $\myo 22.41$\% & $11.30$\% & \myl $16.46$\% & $2.2k/\textrm{class}$ &  \\
        \cline{1-9}
        \myg C\&W & \myo $93.01$\% & $84.98$\% & \myl $53.30$\% & \myo $87.66$\% & $84.16$\% &  \myl $37.76$\% & \myg $2.7k/\textrm{class}$ & \multirow{2}{*}{$\tilde{X}$} \\
        others & \myo $52.54$\% & $80.96$\% & \myl $44.41$\% & \myo $37.12$\% & $81.74$\% & \myl $37.24$\% & $300/\textrm{class}$ & \\
    \hline \hline
        \myg C\&W & \myo $73.03$\% & $15.56$\% & \myl $37.99$\% & \myo $99.96$\% & $15.34$\% & \myl $17.48$\% & \myg $900/\textrm{class}$ & \multirow{2}{*}{$X$} & \multirow{4}{*}{\rotatebox[origin=c]{90}{\parbox[c]{1cm}{\centering Cifar10}}} \\
        others  & \myo $80.07$\% & $16.13$\% & \myl $61.67$\% & \myo $76.13$\% & $11.33$\% & \myl $53.23$\% & $100/\textrm{class}$ & & \\
        \cline{1-9}
        \myg C\&W & \myo $85.33$\% & $95.00$\% & \myl $36.00$\% & \myo $100.00$\% & $95.00$\% & \myl $8.00$\% & \myg $900/\textrm{class}$ & \multirow{2}{*}{$\tilde{X}$} & \\
        others & \myo $82.10$\% & $98.50$\% & \myl $69.93$\% & \myo $78.07$\% & $99.07$\% & \myl $59.80$\% & $100/\textrm{class}$ & & \\
    \bottomrule
    \end{tabular}
\end{table}
}

\section{Introduction}

An adversarial example is a modified input intended to cause a machine-learning model to make a mistake. The modifications are often imperceptible or very subtle to human observers. However, predictive coding can reverse such alterations due to its perturbation resiliency, providing more robustness against such attacks.
This defensive strategy against adversarial attacks includes generative mechanisms that revert the perturbed images to their original form. Predictive coding offers a theoretical framework to support such a defence.

To help understand this work, we will briefly discuss adversarial examples, including how to create and defend against them. We will also mention the attack methods we used in our experiments and popular corresponding defence strategies in subsections \ref{ss:creatingAEs} and \ref{ss:defendingAEs}, respectively. We will introduce the predictive coding framework and its learning algorithm in subsection \ref{ss:PCnet}. We then explain the experiment setups in section \ref{sec:experiments}, and show the results in section \ref{sec:results}, which will be discussed in section \ref{sec:discussion}. At the end, we will summarize our work and point to possible future venues in sections \ref{sec:summary} and \ref{sec:futurework}, respectively.
\subsection{Adversarial Attacks and Defences}\label{ss:introAEs}

Unlike humans, who robustly interpret visual stimuli, artificial neural networks can be deceived by adversarial attacks (ATs), particularly perturbation attacks \cite{kumar2019failure}. These attacks subtly alter an image to trick a well-trained trained feed-forward network (FFnet) used for classification tasks \cite{biggio2013evasion, szegedy2013intriguing} (see figure \ref{fig:advExample}). One standard method to create an adversarial example (AE) that causes the FFnet to misclassify the image as a specific target label is to find a perturbation that minimizes the loss function
\begin{equation}
    \underset{\delta \in \Delta}{\mathrm{argmin}} \ \boldsymbol\ell(F_{\theta}(x+\delta), y_t), \label{eq:loss_AEs_targeted}
\end{equation}
where:
\begin{itemize}
    \item $x$ represents the image,
    \item $\delta$ is the perturbation needed to deceive the FFnet when applied to the image, and $\|\delta\|_{\infty} < \epsilon$ is enforced.
    \item $\Delta$ represents allowable perturbations that are visually indistinguishable to humans.
    \item $y_t$ is the 1-hot vector (i.e., $e_t$) corresponding to the target label $t$.
    \item $F_{\theta}(\cdot)$ is the FFnet model, (i.e., $F_{\theta}: x \rightarrow y \in \mathbb{R}^k$, where $k$ is the number of classes),
    \item $\theta$ represents all parameters defining the model.
    \item $\boldsymbol\ell$ is the cross-entropy loss function.
\end{itemize}
This optimization can be achieved iteratively \cite{carlini2017adversarial}.
Alternatively, instead of deceiving the model by a specific target label, the optimization can be solved for an untargeted attack by maximizing the loss
\begin{equation}
    \underset{\delta \in \Delta(x)}{\mathrm{argmax}} \ \boldsymbol\ell(F_{\theta}(x+\delta), y),
    \label{eq:loss_AEs_untargeted}
\end{equation}
for the given pair $(x, y)$, which can be achieved in one step using methods like the fast gradient sign method (FGSM) \cite{goodfellow2014explaining}, or other approaches \cite{kurakin2018adversarial, madry2017towards, papernot2016limitations, moosavi2016deepfool, moosavi2017universal, xiao2018generating}. When testing a well-trained FFnet MNIST classifier (with an accuracy of approximately $98\%$) against FGSM-generated AEs (with $\epsilon \simeq 0.78\%$), the adversarial success rate is about $41\%$. To defend against ATs, augmenting the training dataset with AEs can improve the classifier's resilience, achieving an accuracy of approximately $94\%$. Alternatively, a min-max approach to directly counteract AEs can enhance robustness within specific perturbation limits \cite{madry2017towards, zhang2019theoretically}.
\figAdvExample
\subsubsection{Creating Adversarial Examples}\label{ss:creatingAEs}

Adversarial examples (AEs) are crucial for assessing and improving the robustness of machine learning (ML) models, especially in deep learning. In image data, creating AEs involves making imperceptible changes to the original image to mislead the model into misclassifying the image. Various methods are available to generate such AEs, particularly for deceiving deep neural networks. Below, we briefly discuss the main gradient-based techniques relevant to our work while noting that there are other techniques to create AEs \cite{papernot2016limitations, moosavi2016deepfool, moosavi2017universal, xiao2018generating}.

\begin{itemize}
    \item Fast Gradient Sign Method (FGSM) modifies the input image by computing the loss gradient for the input image and then making a small step in the opposite direction to increase the loss \cite{goodfellow2014explaining}. 

    \item Basic Iterative Method (BIM), an extension of FGSM, takes multiple small steps while adjusting the direction of the perturbation at each step \cite{kurakin2018adversarial}. 

    \item Projected Gradient Descent (PGD) modifies the input image in multiple iterations with a constraint on the perturbation's size. PGD starts from a random point within a small ball (i.e., $\epsilon$-ball) around the original image and performs a series of gradient descent steps to maximize the prediction error while ensuring the perturbation is smaller than the specified $\epsilon$ \cite{madry2017towards}. 

    \item Carlini \& Wagner (C\&W) attack optimizes the perturbation directly through a loss function that aims to deceive to a desired target label and keep the perturbation small. It often produces subtle perturbations that are highly effective at fooling neural networks \cite{carlini2017adversarial}.
\end{itemize}

These methods differ in complexity, the amount of required knowledge about the target model (white box vs. black box), the type of perturbations (targeted vs. non-targeted), and the strength and stealthiness of the attack. The choice of method often depends on the adversary's access to the model parameters and its specific requirements, including the robustness of the target model and the desired invisibility of the modifications.
\subsubsection{Defending Against Adversarial Attacks}\label{ss:defendingAEs}

Although various adversarial attacks exist, some defence mechanisms attempt to protect ML models against such attacks. Here, we review defence strategies for each previously mentioned attack.

\begin{itemize}
    \item For FGSM and BIM/PGD: Adversarial training involves training the model using adversarial and clean examples. It has been particularly effective against gradient-based attacks like FGSM and BIM. Gradient masking attempts to hide or modify gradients so that they are less useful for generating adversarial examples. However, this method has often been criticized and can be circumvented \cite{papernot2016distillation}.

    \item For C\&W Attack: Some defences estimate the likelihood that input is adversarial using auxiliary models or statistical analyses \cite{carlini2017adversarial}. Defensive distillation involves training a model to output softened probabilities of classes, making it harder for an attacker to find gradients that can effectively manipulate the model’s output \cite{papernot2016distillation}.

\end{itemize}

While these methods offer some protection against specific types of adversarial attacks, it is essential to note that there is no one-size-fits-all solution, and sophisticated or adaptive attackers can circumvent many defences. However, some defence strategies come with a cost, and there is a trade-off between robustness and accuracy \cite{zhang2019theoretically}. Continued research is crucial to improving the robustness of neural networks against these threats.

\subsection{Predictive Coding}\label{ss:PCnet}

Computational neuroscience seeks to understand behavioural and cognitive phenomena at the level of individual neurons or networks of neurons. One approach to solving difficult problems, such as adversarial attacks, which do not seem to be a problem for the brain, is to explore biologically plausible perception models.
The model we will be using is predictive coding (PC)\footnote{Various cortical theories support the bidirectional model \cite{mumford1992computational,rao200216, spratling2010predictive}, as well as free-energy principles \cite{friston2009predictive}.}, a neural model capable of implementing error backpropagation in a biologically plausible manner \cite{bogacz2017tutorial, Millidge2020, Whittington2019}.

\subsubsection{Model Schema and The Learning Algorithms}

The concept of predictive coding suggests that the brain works to minimize prediction error \cite{rao1999predictive}. This model aims to improve overall predictions, and all neurons work towards this common objective. In a predictive coding network (PCnet), each neuron, or \emph{PC unit}, consists of a {\bf value} ($v$) and an {\bf error} node ($\varepsilon$). These PC units are organized into layers, similar to artificial neural networks (ANNs), forming PCnets that learn by adjusting parameters to refine predictions and reduce errors between layers.

For example, in a PCnet, layer $i$ contains vectors $v_{i}$ and $\varepsilon_{i}$, as illustrated in figure \ref{fig:PC}. Vector $v_i$ predicts the values of the next layer, $v_{i-1}$, using prediction weights $M_{i-1}$. The resulting error, $\varepsilon_{i-1}$, is then communicated back via correction weights $W_{i-1}$, allowing $v_i$ to improve its predictions.
\figHierarchicalPC

The network dynamics (as in equations \ref{eq:errorDynamic}~-~\ref{eq:biasDynamic}) are described by the activation function $\sigma$, Hadamard product $\odot$, outer product $\otimes$, decay coefficient $\xi$, and time constants $\tau$ and $\gamma$, where $\tau < \gamma$.
\pcDynamics

where $b_i$ is the bias for error node $\varepsilon_i$. Training a PCnet involves clamping input and output-layer value nodes to sensory input and target values and running the network until it reaches equilibrium. The network's state variables ($v_i$, $\varepsilon_i$) reach equilibrium faster than the parameters ($M_i$, $W_i$, $b_i$) due to $\tau<\gamma$. After training, the parameters $M$, $W$, $b$ are fixed, effectively setting $\gamma$ to infinity.
When a perfect prediction is achieved, the error signal ($\varepsilon$) is zero, stabilizing the value node without further corrections. This state minimizes the Hopfield-like energy function \cite{bogacz2017tutorial} given by equation,
\begin{equation} \label{eq:PC_energy}
    E = \tfrac{\xi}{2} \sum_i \| \vb*{\varepsilon}_i \|^2 .
\end{equation}

After training, when initializing the network with a given input image, when the value nodes are unclamped, the network's ability to reduce energy can lead to potential changes in the input image. The PCnet modifies images without impacting their correct classification, as illustrated in figure~\ref{fig:imageBeforeAfterPC}. The left image displays the original version, while the right image shows the version altered by PCnet. Likewise, when PCnet introduces perturbations to the adversarial image, as seen in figure~\ref{fig:advBeforeAfterPC}, the FFnet can classify it correctly.
\figBeforeAfterPC

PCnet's approach to the credit assignment problem differs from backpropagation (backprop), which is the learning algorithm of ANNs \cite{rumelhart1986learning}. Backpropagation seems unlikely in the brain for several reasons, such as its requirement for weight transposing and transferring between layers. In contrast, PCnet can effectively learn without these requirements using the dynamics of each parameter (equations (\ref{eq:predictionDynamic}), (\ref{eq:correctionDynamic}), and (\ref{eq:biasDynamic})). These dynamics are consistent with the Hebbian learning rule, which only requires local information and pre- and post-synaptic activities \cite{lillicrap2016random, lillicrap2020backpropagation, whittington2017approximation, nokland2016direct, liao2016important, scellier2016towards, hebb2005organization}. This learning algorithm facilitates flexibility and feasibility in using different architectures.
\section{Methodology}\label{sec:experiments}

We conducted experiments to evaluate the effectiveness of our defence strategy against adversarial examples (AEs). We also measured the classification performance of the PCnet and FFnets classifier on AEs. First, we trained the models to classify the MNIST and CIFAR10 datasets. Then, we subjected the FFnets, including a fully connected network (FCnet) and convolutional neural network (CNN), to various adversarial attacks. The experimental setup is illustrated in figure \ref{fig:experiment_diagram}, and the classifiers' architectures are detailed in table \ref{tab:MNISTmodels}. Note that "FFnet" refers to FCnet and CNN in general, and we will repeat the same experiments for both architectures.
\figPCAEsDiagram

To begin with, we partitioned the dataset into \emph{correctly classified} and \emph{misclassified} groups by the PCnet, denoted $X$ and $\tilde{X}$, respectively.\footnote{In this work, we use the term {\bf classified} to indicate trials in which the predicted class matches the correct class. Conversely, we refer to trials where the predicted class differs from the ground truth as {\bf misclassified}.} Next, we randomly chose 20 batches from $X$ and three batches from $\tilde{X}$, each with a batch size of 100. The AT module attacked FFnet and generated AEs $Z$ and $\tilde{Z}$ (corresponds to the given data $X$ and $\tilde{X}$). The attacker ran four methods to create AEs: FGSM, BIM, PGD, and C\&W. Moreover, we impose various $\epsilon$-ball constraints. For the first three methods, the AT module attempted non-targeted attacks and aimed to perturb the image so that FFnet misclassified it. The C\&W method, instead, aimed to produce nine different AEs by targeting all possible labels (0 through 9) for each image.

However, not all AEs are effective. Consequently, we divided these generated AEs into four categories, considering the classifier outcome and the attacker's intention. We depicted successful AEs (misclassified by the FFnet) with shades of red and unsuccessful ones with shades of green, as shown in table~\ref{tab:ABCD}. The first three attack methods are generally non-targeted and fall into categories C and D. The C\&W method aims to create a targeted attack; however, the outcome might not be as intended. In other words, the FFnet might misclassify a created AE to a label other than the intended target (i.e., category C). For example, as in figure~\ref{fig:targeted_non_targeted_AdvBeforePC}, the attacker modifies the image aiming for target labels $8$, $1$, and $3$, from left to right. We call them a failed AE, non-targeted AE, and targeted AE, accordingly. Consequently, the goal is to reduce counts in the A and C categories and increase counts in the B and D categories to enhance the FFnet's robustness.
\tabABCD
\figTargetedNonTargetedAdvBeforePC

In {\bf phase 1}, we fed AEs (e.g., $Z$) into the FFnet and measured the classification accuracy. In {\bf phase 2}, we clamped the input of the PCnet to AEs and measured the classification accuracy. Then, PCnet re-ran to equilibrium, this time \emph{unclamped}, allowing the PCnet to alter the input AEs, which we denoted as $P$ (i.e., $P = \textrm{PCnet}(Z)$). In {\bf phase 3}, we re-evaluate FFnet on the adjusted AEs, $P$. We repeated the same three phases for the inputs that PCnet misclassified in the first place, $\tilde{X}$, and produced $\tilde{Z}$ and $\tilde{P}$, respectively.

We used a GPU machine (AMD FX-8370E, 16GB RAM, and NVIDIA Titan Xp, CUDA 12.1) for the experiment. Approximate timing: training (PCnet 6h, FFnet 5 min), Creating all AEs (C\&W 12h, and FGSM, PGD, BIM $\leq$ 5min), see table \ref{tab:computationalCost} for details.
\tabModelsTwo
\tabComputationalCost
\section{Results}\label{sec:results}

We trained PCnet on a random subset of the MNIST dataset and trained FFnet (FCnet and CNN models) on the entire MNIST dataset. The test accuracy for PCnet, FCnet, and CNN was $74.22\%$, $92.82\%$, and $96.50\%$, respectively. We then partitioned the dataset based on the PCnet classifier: classified ($X$) and misclassified ($\tilde{X}$).
We used the AT module to generate AEs targeting nine incorrect labels per image using the C\&W attack and non-targeted AEs using FGSM, BIM, and PGD attacks. To measure the accuracy, we collected successful AEs that fooled the FFnet and passed them through three phases, as shown in figure~\ref{fig:experiment_diagram}. We conducted the same experiments on the CIFAR10 dataset.

We described the classifier models for the MNIST and CIFAR10 datasets in table~\ref{tab:MNISTmodels}. The ratio of successful AEs relative to the attempted attacks for classified and misclassified data for both sets of experiments on the MNIST and CIFAR10 datasets are shown in table \ref{tab:SuccessfullAttacks} for all three phases: $\textrm{FFnet}(Z)$, $\textrm{PCnet}(Z)$, $\textrm{FFnet}(P)$. 
\tabSuccessfullAttacks

As shown in the figure~\ref{fig:advPredictionCorrectMNIST}, from top to bottom, through these phases: (1) we cherry-picked AEs, $Z$, that were $100\%$ successful against FCnet; (2) PCnet classified $70\%$ of those AEs, and adjusted $Z$ into $P$ in unclamped mode; (3) this time, FCnet classified $81\%$ of adjusted AEs, $P$. Similarly, on $100\%$ successful AEs attacking the CNN model, PCnet classified $68\%$ of them, and the CNN model classified $84\%$ of adjusted such AEs.
On average, the number of successful AEs per class for FCnet is higher than for the CNN model, as the FCnet is simpler and easier to fool. At the same time, the CNN model benefits more from adjusted AEs than FCnet. Similar trends were observed for the CIFAR10 dataset, as shown in figure ~\ref{fig:advPredictionCorrectCIFAR}, where the CNN model outperformed the FCnet.

We also applied the same AT methods to generate $\tilde{Z}$: AEs from the misclassified subset $\tilde{X}$ on attacking FFnet. Then we passed $\tilde{Z}$ through our multi-phase experiment (as shown in figure~\ref{fig:experiment_diagram}).
As PCnet misclassified $\tilde{X}$ in the first place, we did not expect PCnet to perform well on corresponding AEs. However, the results showed that perturbing the AEs affected PCnet's classification accuracy, which might be the result of two consequence perturbations: one made by an attacker to craft AEs and another one when PCnet adjusts AEs seeking a lower based on its dynamics. However, as the primary goal, the FFnet models performed better on adjusted AEs compared to the raw AEs, with an increase in classification accuracy by $44\%$ for FCnet and $57\%$ for CNN models, as shown in figure \ref{fig:advPredictionWrongMNIST}.
\figAdvPredictionCorrectMNIST
\figAdvPredictionCorrectCIFAR
\figAdvPredictionWrongMNIST
\section{Discussion}\label{sec:discussion}

Now, we are going to scrutinize the results.
When creating Adversarial Examples (AEs) from a random subset of each class, some classes are more vulnerable than others. Moreover, the vulnerability level for each class might not be the same across different models. For example, as shown in figures~\ref{fig:advPredictionPC&FC}, \ref{fig:advPredictionPC&CNN}, about $90\%$ of attacks on FCnet and CNN for class `$1$' were successful, while the numbers for class `$0$' are less than $60\%$ and $30\%$, respectively. However, on average, FCnet is more deceivable than CNN.
Nevertheless, PCnet can classify many AEs created to deceive either model, regardless of targeted or non-targeted attacks (as shown in figure~\ref{fig:advPredictionCorrectMNIST}). 

Figure \ref{fig:advPredictionPC&FC} quantifies classification accuracy on AEs for FCnet on different classes. In some classes, like `$6$', the performances of PCnet on AEs and FFnet on adjusted AEs are similar. Likewise, in classes `$6$' and `$7$', the performances of PCnet on AEs and CNN on adjusted AEs are similar, as shown in figure~\ref{fig:advPredictionPC&CNN}.

\figAdvPredictionsFC
\figAdvPredictionsCNN

Moreover, we observed that PCnet is less skillful in classifying digit `$5$' than other classes. Consequently, PCnet's classification ability on AEs for class `$5$' is limited compared to AEs from others. However, the PCnet reversion process on AEs improves the classification of `$5$' for FFnet. On average, FFnet shows better classification when exposed to adjusted AEs by PCnet (i.e., $P$) than PCnet on AEs (i.e., $Z$). As the accuracy of FFnet models is better than that of PCnet in the first place, FFnet classification on adjusted AEs is better, too. In other words, we can improve the lower bound and close the gap by improving the accuracy of PCnet.

Partitioning AEs into the categories explained in the table~\ref{tab:ABCD}, the classification accuracy of FFnet improved over the process of modifying $Z$ into $P$, as shown in figure~\ref{fig:advPredictionPC&FFnet}. Besides, the CNN classification on adjusted AEs, $P$, shows collectively better performance on different classes (i.e., a lower variance), which is consistent with $6\%$ misclassification rate for CNN in figure~\ref{fig:advPredictionCorrectMNIST}.
\figAdvPredictionsFFnet

What modifications does PCnet make on AEs? When we clamp the input of PCnet with an AE, the network's energy approaches an equilibrium (as shown in the first half of figure \ref{fig:advPCenergy}). By clamping the PCnet input to the AE, we create tension in the network because the input does not fall within the generative range of the network. However, unclamping the input releases the stress, and the network's energy decreases as its state moves toward equilibrium, as shown in the second half of figure~\ref{fig:advPCenergy}. As a result, the PCnet alters the AE and converges to an equilibrium consistent with its generative expectation.
\figAdvPCEnergy

When we perturb an image to create AEs using gradient ascent, we can move away from a local minimum, for example, from point $O_1$ to point $T_1$ as shown in figure~\ref{fig:perturb_reverte}. On the other hand, the dynamical structure of PCnet guides the network's state towards lower energy, reaching equilibrium, i.e., from point $T_1$ back to point $O_1$ in figure~\ref{fig:perturb_reverte}. At point $T_1$, the FFnet misclassifies the AE, as it falls into the other side of the decision boundary but still in the basin of attraction $O_1$ compared to others.
If we modify the image so that the AE falls into the basin of another attractor (crosses a PCnet decision boundary), the PC dynamics will not bring it back to the original basin, but instead push the AE further toward a different equilibrium. In this case, both the FFnet and PCnet would misclassify the adjusted AE. For instance, as illustrated in figure~\ref{fig:perturb_reverte}, when attempting to create an AE from $O_2$, the point $T_2$ crossed the FFnet decision boundary and ended up in the basin of $O_3$ instead of $O_2$. Consequently, when the PCnet further modifies AE, it moves even closer to $O_3$.
\figPerturbRevert

However, improving the classification of adjusted AEs comes with a cost. The relocation between decision boundaries and PCnet adjustments is not always favourable for improving the classification. Based on the categories depicted in table \ref{tab:ABCD}, we aim to convert more misclassified instances (i.e., categories A and C) into classified instances (i.e., categories B and D) after our process. However, in some cases, we observed that some AEs previously classified turned into misclassified ones after PCnet's modifications. Overall, as depicted in figures \ref{fig:AE_FCnet_classifiedMNIST} and \ref{fig:AE_CNN_classifiedMNIST}, the amount of improvement (depicted in the green bands) outweighed the cost of loss or failure (depicted in red bands).
\figSankeyFCnetClassifiedMNIST
\figSankeyCNNClassifiedMNIST

We noticed that PCnet modifications improved FFnet performance on AEs created from instances classified by PCnet, i.e., $X$, as shown in figures \ref{fig:AE_FCnet_classifiedMNIST} and \ref{fig:AE_CNN_classifiedMNIST}. Furthermore, PCnet dynamics, which modifies AEs from misclassified instances, i.e., $\tilde{X}$, helped FFnet to enhance its performance, as shown in figures \ref{fig:AE_FCnet_misclassifiedMNIST} and \ref{fig:AE_CNN_misclassifiedMNIST}. It is important to note that the perturbations caused by the adversarial attacks, in some cases, pushed the data point toward the correct basin of attraction. Consequently, the PCnet classified the AE, even though the PCnet misclassified it before perturbation.
\figSankeyFCnetmisClassifiedMNIST
\figSankeyCNNmisClassifiedMNIST

We created adversarial examples (AEs) using four different methods: FGSM, BIM, PGD, and C\&W. The C\&W method is more complex and time-consuming for developing AEs through targeted attacks. As shown in table \ref{tab:SuccessfullAttacks}, the PCnet reversion process is effective with the C\&W method. We applied the first three methods with varying $\epsilon$-balls; however, they all exhibited similar behaviour. The results presented in table \ref{tab:SuccessfullAttacks} is the average of three methods (for $\epsilon=0.75$), which behave similarly since they are variations of FGSM.

Table \ref{tab:SuccessfullAttacks} indicates a consistent improvement in FFnet's classification when using perturbed AEs. In other words, the perturbation applied by PCnet significantly enhances the FFnet's classification accuracy by a minimum factor of two. Note that, the perturbation remains beneficial for classifying AEs built on unfamiliar samples $\tilde{X}$.

Furthermore, since the target is not enforced for FGSM, PGD, and BIM methods, AEs might converge to a state misclassified by PCnet—in other words, crossing both the PCnet and FFnet decision boundaries. In contrast, a targeted attack iteratively tries to cross a specific boundary.

To better grasp the perturbation and reversion process, we created an AE by attacking the FFnet (e.g., FCnet) using the image of $0$ and targeting label $3$. As in figure~\ref{fig:FC_adjusted_adv}, the red arrow shows how rapidly the FFnet loses confidence on $0$ and is fooled and misled to $3$ as we perturb the image. Conversely, the green arrow shows how the changes in PCnet's state revert the attack process, where FFnet gains confidence for $0$ after PCnet modifications in several steps.
\figPredcitxHat

As CNN specializes in extracting features in several layers and combining them later to perform the classification tasks, the PCnet modification benefits CNN more. Here, we applied different filters on AEs to see if the FFnet performs similarly to when they get modified AEs from PCnet. As in figure~\ref{fig:PCnet_vs_filters}, the original image and the corresponding AE are followed by $P_\mathrm{cut off}$ and $p$, then a series of filtered ones. $P_\mathrm{cut off}$ is the instance that just passed the decision boundary but not reached equilibrium $p$. Their probabilities show that PCnet modification is more effective than others. For example, consider the Gaussian filter applied on AE to become $g$ (i.e., $g = \textrm{GaussianBlur}(x)$). If we plot mappings from AE to the image $x$, we can find that $\cos(p, x) > \cos(g, x)$. Moreover, when PCnet modifies $x$ into $\hat{x}$, it becomes very similar to $p$. This hints at why PCnet can classify AEs well, as shown in figure \ref{fig:PCnet_vs_filters_cosine}.
\figPCvsFilters
\figPCvsFiltersCosine

We compared the performance of using an image filter against using the PCnet. Instead of the PCnet modifications on AEs, we applied the \emph{GaussianBlur} filter on AEs. The figures \ref{fig:PCnet_filter_boxplot_MNIST_CIFAR10_FC_correct}, \ref{fig:PCnet_filter_boxplot_MNIST_CIFAR10_CNN_correct}, \ref{fig:PCnet_filter_boxplot_MNIST_CIFAR10_FC_wrong}, \ref{fig:PCnet_filter_boxplot_MNIST_CIFAR10_CNN_wrong} show that the PCnet defence strategy is more effective in improving the classification performance of FFnets, while the performance of FFnet on modification through the filter lags behind.
\figPCvsFiltersCorrectFCBoxplotMNISTCIFAR
\figPCvsFiltersCorrectCNNBoxplotMNISTCIFAR
\figPCvsFiltersWrongFCBoxplotMNISTCIFAR
\figPCvsFiltersWrongCNNBoxplotMNISTCIFAR

\subsection{Related Works}

In a related work, the pre-trained FFnet is augmented with PCnet. At each layer of the FFnet, the generative feedback predicts the pattern of activities in the previous layer. Thus, the reconstruction errors iteratively help to improve the network's representation \cite{choksi2020brain}. In other words, PCnet attaches to FFnet to compensate for its lack of dynamics. Contrary to our strategy, in this model, we should know the architecture of the underlying model.
However, in our defence strategy, PCnet is unaware of the FFnet model or the training process for this approach. Instead, PCnet uses its dynamic properties to modify the image to settle in a lower level of network energy. In contrast, FFnet lacks a dynamic mechanism to alter or adjust any possible perturbation. 

In other words, by clamping PCnet to the input, the network runs to equilibrium and shows a stable energy level. However, the network's state changes once we release the clamping restriction, seeking a lower energy level. So, when PCnet pushes the image toward its local minima, the modified image should fall into the correct FFnet decision boundary as long as PCnet classifies it correctly and makes sense of it. For this purpose, PCnet needs to be trained on the same dataset, which is the only matter that PCnet and FFnet have in common.
\section{Summary}\label{sec:summary}

Our observations on both correctly classified $X$ and misclassified $\Tilde{X}$ data (as shown in figures \ref{fig:advPredictionCorrectMNIST}, \ref{fig:advPredictionWrongMNIST}) suggest that using the PCnet reversion process benefits FFnet. These results hold not only for the inputs that the PCnet correctly classified but also for the inputs that the PCnet misclassified. 

The PCnet's generative nature results in lower classification accuracy. Improving this accuracy would presumably further improve the robustness of the PCnet reversion process, which means more AEs can be reverted.

The PCnet performs well in classifying AEs, but its primary purpose is to enhance the classification performance of FFnets without directly accessing them. The modifications made to the PCnet serve as a preprocessing step, ultimately improving the classification accuracy of FFnets.

Using PCnet as a pre-processor for FFnets presents a promising defence strategy against adversarial attacks on neural network classifiers. Guided by generative functionality, PCnets effectively revert adversarial perturbations, pushing images closer to their original forms. While demonstrating efficacy on the MNIST and CIFAR10 datasets with FCnet and CNN architectures, further research is needed to assess scalability, generalization, and robustness across diverse datasets and adversarial attack scenarios. Nonetheless, PCnet offers a significant step towards enhancing the security and reliability of neural network classifiers in the face of evolving adversarial threats.

\section{Future Work}\label{sec:futurework}

The study highlights the effectiveness of PCnets in defending against adversarial attacks, but there are potential gaps to consider.
\begin{enumerate}
    \item {\bf Scalability}: We conducted the experiments on the MNIST and CIFAR10 datasets, which are relatively simple compared to some high-dimensional real-world datasets. Further research is needed to assess the scalability of PCnets to larger and more complex datasets.
    
    \item {\bf Generalization}: The study focuses on specific architectures (FC and CNN) and datasets. However, to ensure their applicability in diverse scenarios, it is essential to investigate the generalization of PCnets across various network architectures and datasets.
    
    \item {\bf Adversarial Strength}: The study focuses on four major gradient-based adversarial attacks. Further exploration is needed to assess the effectiveness of PCnets against more sophisticated adversarial attacks.
    
    \item {\bf Complex PCnet models}: For our experiments, we constructed PCnets in a fully connected architecture and trained them by clamping the training dataset without using any specific regularization technique. However, a more complex architecture on PCnet might excel in defending against more complex attacks or datasets.

    \item {\bf Computational Overhead}: We trained PCnets using only $10\%$ random subset of each dataset; however, training PCnets was a bottleneck in our experiment pipelines, so the computational cost of PCnet as a defence mechanism needs to be evaluated, particularly in real-time or resource-constrained environments.

    \item {\bf Ineffectiveness}: PCnet effectively defends against C\&W attacks. However, more analytical experiments are needed to show why it is ineffective against FGSM, PGD, and BIM attacks.
\end{enumerate}

Addressing these gaps will further validate the feasibility and effectiveness of PCnets as a defence mechanism against adversarial attacks on neural network classifiers.

\clearpage

\begin{thebibliography}{10}

\bibitem{kumar2019failure}
Ram Shankar~Siva Kumar, David~O Brien, Kendra Albert, Salom{\'e} Vilj{\"o}en, and Jeffrey Snover.
\newblock Failure modes in machine learning systems.
\newblock {\em arXiv preprint arXiv:1911.11034}, 2019.

\bibitem{biggio2013evasion}
Battista Biggio, Igino Corona, Davide Maiorca, Blaine Nelson, Nedim {\v{S}}rndi{\'c}, Pavel Laskov, Giorgio Giacinto, and Fabio Roli.
\newblock Evasion attacks against machine learning at test time.
\newblock In {\em Machine Learning and Knowledge Discovery in Databases: European Conference, ECML PKDD 2013, Prague, Czech Republic, September 23-27, 2013, Proceedings, Part III 13}, pages 387--402. Springer, 2013.

\bibitem{szegedy2013intriguing}
Christian Szegedy, Wojciech Zaremba, Ilya Sutskever, Joan Bruna, Dumitru Erhan, Ian Goodfellow, and Rob Fergus.
\newblock Intriguing properties of neural networks.
\newblock {\em arXiv preprint arXiv:1312.6199}, 2013.

\bibitem{carlini2017adversarial}
Nicholas Carlini and David Wagner.
\newblock Adversarial examples are not easily detected: Bypassing ten detection methods.
\newblock In {\em Proceedings of the 10th ACM workshop on artificial intelligence and security}, pages 3--14, 2017.

\bibitem{goodfellow2014explaining}
Ian~J Goodfellow, Jonathon Shlens, and Christian Szegedy.
\newblock Explaining and harnessing adversarial examples.
\newblock {\em arXiv preprint arXiv:1412.6572}, 2014.

\bibitem{kurakin2018adversarial}
Alexey Kurakin, Ian~J Goodfellow, and Samy Bengio.
\newblock Adversarial examples in the physical world.
\newblock In {\em Artificial intelligence safety and security}, pages 99--112. Chapman and Hall/CRC, 2018.

\bibitem{madry2017towards}
Aleksander Madry, Aleksandar Makelov, Ludwig Schmidt, Dimitris Tsipras, and Adrian Vladu.
\newblock Towards deep learning models resistant to adversarial attacks.
\newblock {\em arXiv preprint arXiv:1706.06083}, 2017.

\bibitem{papernot2016limitations}
Nicolas Papernot, Patrick McDaniel, Somesh Jha, Matt Fredrikson, Z~Berkay Celik, and Ananthram Swami.
\newblock The limitations of deep learning in adversarial settings.
\newblock In {\em 2016 IEEE European symposium on security and privacy (EuroS\&P)}, pages 372--387. IEEE, 2016.

\bibitem{moosavi2016deepfool}
Seyed-Mohsen Moosavi-Dezfooli, Alhussein Fawzi, and Pascal Frossard.
\newblock Deepfool: a simple and accurate method to fool deep neural networks.
\newblock In {\em Proceedings of the IEEE conference on computer vision and pattern recognition}, pages 2574--2582, 2016.

\bibitem{moosavi2017universal}
Seyed-Mohsen Moosavi-Dezfooli, Alhussein Fawzi, Omar Fawzi, and Pascal Frossard.
\newblock Universal adversarial perturbations.
\newblock In {\em Proceedings of the IEEE conference on computer vision and pattern recognition}, pages 1765--1773, 2017.

\bibitem{xiao2018generating}
Chaowei Xiao, Bo~Li, Jun-Yan Zhu, Warren He, Mingyan Liu, and Dawn Song.
\newblock Generating adversarial examples with adversarial networks.
\newblock {\em arXiv preprint arXiv:1801.02610}, 2018.

\bibitem{zhang2019theoretically}
Hongyang Zhang, Yaodong Yu, Jiantao Jiao, Eric Xing, Laurent El~Ghaoui, and Michael Jordan.
\newblock Theoretically principled trade-off between robustness and accuracy.
\newblock In {\em International conference on machine learning}, pages 7472--7482. PMLR, 2019.

\bibitem{papernot2016distillation}
Nicolas Papernot, Patrick McDaniel, Xi~Wu, Somesh Jha, and Ananthram Swami.
\newblock Distillation as a defense to adversarial perturbations against deep neural networks.
\newblock In {\em 2016 IEEE symposium on security and privacy (SP)}, pages 582--597. IEEE, 2016.

\bibitem{mumford1992computational}
David Mumford.
\newblock On the computational architecture of the neocortex: Ii the role of cortico-cortical loops.
\newblock {\em Biological cybernetics}, 66(3):241--251, 1992.

\bibitem{rao200216}
Rajesh~PN Rao and Terrence~J Sejnowski.
\newblock Predictive coding, cortical feedback, and spike-timing dependent plasticity.
\newblock {\em Probabilistic models of the brain}, page 297, 2002.

\bibitem{spratling2010predictive}
Michael~W Spratling.
\newblock Predictive coding as a model of response properties in cortical area {V1}.
\newblock {\em Journal of Neuroscience}, 30(9):3531--3543, 2010.

\bibitem{friston2009predictive}
Karl Friston and Stefan Kiebel.
\newblock Predictive coding under the free-energy principle.
\newblock {\em Philosophical transactions of the Royal Society B: Biological sciences}, 364(1521):1211--1221, 2009.

\bibitem{bogacz2017tutorial}
Rafal Bogacz.
\newblock A tutorial on the free-energy framework for modelling perception and learning.
\newblock {\em Journal of mathematical psychology}, 76:198--211, 2017.

\bibitem{Millidge2020}
Beren Millidge, Alexander Tschantz, and Christopher~L. Buckley.
\newblock {Predictive Coding Approximates Backprop along Arbitrary Computation Graphs}.
\newblock {\em arXiv}, pages 1--27, 2020.

\bibitem{Whittington2019}
James C.~R. Whittington and Rafal Bogacz.
\newblock {Theories of Error Back-Propagation in the Brain}.
\newblock {\em Trends in Cognitive Sciences}, 23(3):235--250, 2019.

\bibitem{rao1999predictive}
Rajesh~PN Rao and Dana~H Ballard.
\newblock Predictive coding in the visual cortex: a functional interpretation of some extra-classical receptive-field effects.
\newblock {\em Nature Neuroscience}, 2(1):79, 1999.

\bibitem{rumelhart1986learning}
David~E Rumelhart, Geoffrey~E Hinton, and Ronald~J Williams.
\newblock Learning representations by back-propagating errors.
\newblock {\em Nature}, 323(6088):533, 1986.

\bibitem{lillicrap2016random}
Timothy~P Lillicrap, Daniel Cownden, Douglas~B Tweed, and Colin~J Akerman.
\newblock Random synaptic feedback weights support error backpropagation for deep learning.
\newblock {\em Nature Communications}, 7:13276, 2016.

\bibitem{lillicrap2020backpropagation}
Timothy~P Lillicrap, Adam Santoro, Luke Marris, Colin~J Akerman, and Geoffrey Hinton.
\newblock Backpropagation and the brain.
\newblock {\em Nature Reviews Neuroscience}, 21(6):335--346, 2020.

\bibitem{whittington2017approximation}
James~CR Whittington and Rafal Bogacz.
\newblock An approximation of the error backpropagation algorithm in a predictive coding network with local hebbian synaptic plasticity.
\newblock {\em Neural computation}, 29(5):1229--1262, 2017.

\bibitem{nokland2016direct}
Arild N{\o}kland.
\newblock Direct feedback alignment provides learning in deep neural networks.
\newblock {\em Advances in neural information processing systems}, 29, 2016.

\bibitem{liao2016important}
Qianli Liao, Joel~Z Leibo, and Tomaso~A Poggio.
\newblock How important is weight symmetry in backpropagation?
\newblock In {\em AAAI}, pages 1837--1844, 2016.

\bibitem{scellier2016towards}
Benjamin Scellier and Yoshua Bengio.
\newblock Towards a biologically plausible backprop.
\newblock {\em arXiv preprint arXiv:1602.05179}, 914, 2016.

\bibitem{hebb2005organization}
Donald~Olding Hebb.
\newblock {\em The organization of behavior: A neuropsychological theory}.
\newblock Psychology press, 2005.

\bibitem{choksi2020brain}
Bhavin Choksi, Milad Mozafari, Callum~Biggs O'May, Benjamin Ador, Andrea Alamia, and Rufin VanRullen.
\newblock Brain-inspired predictive coding dynamics improve the robustness of deep neural networks.
\newblock In {\em Neurips 2020 workshop svrhm}, 2020.

\end{thebibliography}

\end{document}